\documentclass[11pt]{article}
\usepackage{amstext, amsmath,latexsym,amsbsy,amssymb,amsmath}

\usepackage{enumitem}

\newtheorem{remark}{Remark}[section]

\numberwithin{equation}{section}

\newcommand{\QED}{\hspace*{\fill}\rule{2.5mm}{2.5mm}}

\usepackage{amssymb}\usepackage{graphicx}
\usepackage{tikz}

\usepackage{color}
\newcommand\qed{\hfill$\sqcap\kern-7.5pt\hbox{$\sqcup$}$}

\newcommand{\beqn}{\begin{equation}}
\newcommand{\eeqn}{\end{equation}}
\newcommand{\bear}{\begin{eqnarray}}
\newcommand{\eear}{\end{eqnarray}}
\newcommand{\bean}{\begin{eqnarray*}}
\newcommand{\eean}{\end{eqnarray*}}

\allowdisplaybreaks

\begin{document}
\title{Quantum hydrodynamic approximations to the finite temperature trapped Bose gases}

\author{Shi Jin\footnotemark[1] \and Minh-Binh Tran\footnotemark[2] 
}

\renewcommand{\thefootnote}{\fnsymbol{footnote}}

\footnotetext[1]{Department of Mathematics, University of Wisconsin-Madison, Madison, WI 53706, USA. \\Email: sjin@wisc.edu
}

\footnotetext[2]{Department of Mathematics, University of Wisconsin-Madison, Madison, WI 53706, USA. \\Email: mtran23@wisc.edu
}

\maketitle
\begin{abstract} 
For the quantum kinetic system modelling the Bose-Einstein Condensate that
accounts for interactions between condensate and excited atoms, we
use the Chapman-Enskog expansion to derive its hydrodynamic approximations,
include both Euler and Navier-Stokes approximations. The hydrodynamic approximations describe not only the macroscopic behavior of  the BEC but also its coupling with the
non-condensates, which agrees with the Landau two-fluid theory.

 \end{abstract}

{\bf Keyword:}
{Low and high temperature quantum kinetics; Bose-Einstein  condensate; quantum Boltzmann equation;  defocusing cubic nonlinear Schrodinger equation; quantum hydrodynamics limit. 


\tableofcontents
\section{Introduction}
After the realization of Bose-Einstein condensations (BECs) in trapped atomic vapors of $^{87}$Rb, $^7$Li, and $^{23}$Na \cite{WiemanCornell,Ketterle}, a new period of intense experimental and theoretical research has been initiated.  The equilibrium properties of these novel systems have been quite well understood, but there are still several open questions concerning their nonequilibrium behavior. One of the most important questions  concerns the behavior of the condensate after cooling a nondegenerate trapped Bose gas to a  temperature below the BEC critical temperature. While the experimental research  has, up to now, concentrated mainly on the initial formation of BECs, their theoretical behaviour at finite temperatures is a frontier of many-body physics.  The theoretical description of BECs has to take into account the coupled nonequilibrium dynamics of both the condensed and  noncondensed components of  the  gas under investigation, and has to involve the collisional processes of atoms between the two components. Such a quantum kinetic theory was inititated by    Kirkpatrick and Dorfman \cite{KD1,KD2}, based on the rich body of research carried out in the period 1940-67 by Bogoliubov, Lee and Yang, Beliaev, Pitaevskii, Hugenholtz and Pines, Hohenberg and Martin, Gavoret and Nozi`eres, Kane and Kadanoff and many others. The terminology ``Quantum Kinetic Theory'' has been later introduced in a series of papers by Gardinier, Zoller and collaborators  \cite{QK0,QK1,QK2,QK3}. After that, there has been an explosion of research on  quantum kinetic theory  (see \cite{josserand2001nonlinear,BaoCai, BPM, bijlsma2000condensate,KD1,KD2,Spohn:2010:KOT,GriffinNikuniZaremba:2009:BCG,QK0,QK1,QK2,QK3,ArkerydNouri:2012:BCI,ArkerydNouri:2015:BCI,ArkerydNouri:AMP:2013,ReichlGust:2013:RRA,pomeau1999theorie,HuJin:2011:OKF,FilbetHuJin:2012:ANS,DyachenkoNewellPushkarev:OTW:1992,ZakharovNazarenko:DOT:2005,zaremba1998two}, and references therein).  We refer to the review paper \cite{anglin2002bose} and the books \cite{inguscio1999bose,ColdAtoms1}, for more discussions and a complete list of references on this rapidly expanding topic. 

The current paper is devoted to the study of the hydrodynamic approximations of such a quantum kinetic system. The system contains two equations: a quantum Boltzmann equation describing the non-condensate atoms (with two types of collisions, one between excited atoms and one between condensate atoms and excited atoms), and a nonlinear Schr\"odinger (or Gross-Pitaevski) equation for the condensate. The hydrodynamic limits of the system is an interesting mathematical question, first studied in \cite{Allemand:Thesis:2009}, where  an Euler limit  has been derived.  This derivation relies on the assumption that,  in the considered trapped Bose gas, the noncondensate and condensate share the same 
local equilibrium. It is known (cf. \cite{KD1,KD2}) that the condition of
complete local equilibrium between the condensate and the thermal cloud
requires the energy of a condensate atom in the
local rest frame of the thermal cloud to be equal to the local thermal cloud chemical
potential. When the condition is satisfied, there is no exchange of particles
between the condensate and the thermal cloud (cf. \cite{GriffinNikuniZaremba:2009:BCG}). As a consequence, in  the derived fluid system, the mass of each component - condensate and non-condensate -  does not exchange. Note that the  two-fluid low-frequency dynamics of superfluid $^4$He was first developed
by Tisza and Landau \cite{lifshitz1987fluid}. Their description accounts for the characteristic features associated with
superfluidity in terms of the relative motion of superfluid
and normal fluid degrees of freedom, and was shown  to be a consequence of a Bose broken symmetry (cf.  \cite{bogoliubov1970lectures}). In the Landau two-fluid theory, the two components superfluid
and normal fluid  exchange mass (cf. \cite{Allemand:Thesis:2009,lifshitz1987fluid,bogoliubov1970lectures}). In this paper, we revisit the derivation of the Euler hydrodynamic limit of the system by a different point of view: following \cite{KD1,KD2,GriffinNikuniZaremba:2009:BCG}, we assume that  even if the  thermal cloud atoms are in equilibrium among
themselves, the noncondensate and condensate
parts may not be in local equilibrium with each other.  Moreover, the derivation of the Navier-Stokes approximation of the system is also provided via the
classical Chapman-Enskog expansion (cf. \cite{TruesdellMuncaster:FOM:1980}). In
such circumstance, the Euler limit includes the mass exchange between the
condensate and the non-condensate.  Our Euler and Navier-Stokes approximations
agree with the Landau two-fluid theory (cf. \cite{lifshitz1987fluid,bogoliubov1970lectures}).

 As an attempt to build a rigourous theory for quantum kinetic equations, some mathematical results have been obtained in \cite{AlonsoGambaBinh,CraciunBinh,Binh9,GambaSmithBinh,GermainIonescuTran,ToanBinh,nguyen2017quantum,SofferBinh1,ReichlTran,SofferBinh2})  . Note that quantum kinetic equations have very similar formulations with the so-called wave turbulence kinetic equations. We refer to \cite{buckmaster2016analysis,FaouGermainHani:TWN:2016,germain2015high,germain2015continuous,LukkarinenSpohn:WNS:2011,Nazarenko:2011:WT,Spohn:WNW:2010,zakharov2012kolmogorov,Zakharov:1998:NWA} for more recent advances on the rigorous theory of weak turbulence.

The plan of the paper is as follows. In Section \ref{Sec2} we introduce
the quantum kinetic system and the scalings that will lead to the 
hydrodynamic approximation. 
In Section \ref{Sec:Operators}, we list the most important features of the two collision operator $C_{12}$ and $C_{22}$.
The two-fluid Euler and  Navier-Stokes limits are then derived in the two Sections \ref{Sec:Euler} and \ref{Sec:NavierStokes} respectively.

\section{The quantum kinetic system and scalings}\label{Sec2}
%
%
%
%

\subsection{The quantum kinetic system}

Let us consider a trap Bose gas, whose temperature $T$ is smaller than the Bose-Einstein transition temperature $T_{BEC}$ and strictly greater than $0$  K or $-273.15^oC$. Denote $f(t,r,p)$ to be the density function of the normal fluid at time $t$, position $r$ and momentum $p$ and $\Phi(t,r)$ be the wave function of the  the condensated (or superfluid) phase. Employing the short-handed notation
$f_i=f(t,r,p_i)$, $i=1,2,3,4$,
we first recall the  quantum kinetic - Schr\"odinger system describing the dynamics of a BEC and its thermal cloud. The Schr\"odinger (or the Gross-Pitaevski) equation for the condensates reads (cf. \cite{bijlsma2000condensate}):
\begin{equation}
\begin{aligned}\label{GP}
i \hbar {\partial_t \Phi(t,r)} =&\ \Big(-\frac{\hbar^2 \Delta_{{r}}}{2m}+g[n_c(t,r) + 2n_n(t,r)] -i\Lambda_{12}[f](t,r) +V(r)\Big)\Phi(t,r), \ \ (t,r)\in\mathbb{R}_+\times\mathbb{R}^3,\\
\Lambda_{12}[f](t,r) = &\ \frac{\hbar}{2n_c}\Gamma_{12}[f](t,r),\\
\Gamma_{12}[f](t,r)= &\ \int_{\mathbb{R}^{3}}C_{12}[f](t,r,p)\frac{d p}{(2\pi \hbar)^3},\\ 
 n_n(t,r) \ = & \ \int_{\mathbb{R}^3}f(t,r,p)d p,
 \\
~~~\Phi(0,r)=& \ \Phi_0(r), \forall r\in\mathbb{R}^3,
\end{aligned}
\end{equation}
where $n_c(t,r)=|\Phi|^2(t,r)$ is the condensate density, $\hbar$ is the Planck constant, $g$ is the interaction coupling constant proportional to  the $s$-wave scattering length $a$, $V(r)$ is the confinement potential, and the operator $C_{12}$ can be found in   
the quantum Boltzmann equation for the non-condensate atoms (cf. \cite{bijlsma2000condensate}), written below:
\begin{eqnarray}\label{QBFull}
{\partial_t f}(t,r,p)&+&\frac{p}{m}\cdot\nabla_{{r}} f(t,r,p) \ - \ \nabla_r U(t,r)\cdot \nabla_p f(t,r,p)\\\nonumber
&=&Q[f](t,r,p):=C_{12}[f](t,r,p)+C_{22}[f](t,r,p), (t,r,p)\in\mathbb{R}_+\times\mathbb{R}^3\times\mathbb{R}^3,\\\nonumber
C_{12}[f](t,r,p_1)&:=&\lambda_1n_c(t,r) \iint_{\mathbb{R}^{3}\times\mathbb{R}^{3}}\delta(mv_c+ {p}_1-{p}_2-{p}_3)\delta(\mathcal{E}_c+\mathcal{E}_{{p}_1}-\mathcal{E}_{{p}_2}-\mathcal{E}_{{p}_3})\\\label{C12}
& &\times[(1+f_1)f_2f_3-f_1(1+f_2)(1+f_3)]d p_2d p_3\\\nonumber
&&-2\lambda_1n_c(t,r)\iint_{\mathbb{R}^{3}\times\mathbb{R}^{3}}\delta(mv_c+{p}_2-{p}_1-{p}_3)\delta(\mathcal{E}_c+\mathcal{E}_{{p}_2}-\mathcal{E}_{{p}_1}-\mathcal{E}_{{p}_3})\\\nonumber
& &\times[(1+f_2)f_1f_3-f_2(1+f_1)(1+f_3)]d p_2d p_3,\\\label{C22}
C_{22}[f](t,r,p_1)&:=&\lambda_2\iiint_{\mathbb{R}^{3}\times\mathbb{R}^{3}\times\mathbb{R}^{3}}\delta({p}_1+{p}_2-{p}_3-{p}_4)\\\nonumber
& &\times\delta(\mathcal{E}_{{p}_1}+\mathcal{E}_{{p}_2}-\mathcal{E}_{{p}_3}-\mathcal{E}_{{p}_4})\times\\\nonumber
&&\times [(1+f_1)(1+f_2)f_3f_4-f_1f_2(1+f_3)(1+f_4)]d p_2d p_3d p_4,
\\\nonumber
f(0,r,p)&=&f_0(r,p), (r,p)\in\mathbb{R}^3\times\mathbb{R}^3,
\end{eqnarray}
where $\lambda_1=\frac{2g^2}{(2\pi)^2\hbar^4},$ $\lambda_2=\frac{2g^2}{(2\pi)^5\hbar^7}$, $m$ is the mass of the particles, $\mathcal{E}_{{p}}$ is the Hartree-Fock  energy (cf. \cite{bijlsma2000condensate})
\bear \label{def-E}
\mathcal{E}_p\ =\ \mathcal{E}(p)\ = \ \frac{|p|^2}{2m} + U(t,r).
\eear
Notice that $C_{22}$ is the Boltzmann-Norheim (Uehling-Ulenbeck) quantum Boltzmann collision operator. 
If one writes  
\begin{equation}\label{def-Phi}
\begin{aligned}
\Phi \ = & \ |\Phi(t,r)|e^{i\phi(t,r)},
\end{aligned}
\end{equation}
the condensate velocity can be defined as
\begin{equation}\label{def-vc}
v_c(t,r) = \frac{\hbar}{m}\nabla \phi(t,r),
\end{equation}
and the condensate chemical potential is then
\begin{equation}\label{def-muc}
 \mu_c 
=\ \frac{1}{\sqrt{n_c}}\left(-\frac{\hbar^2\Delta_r}{2m}+ V +g[2n_n+n_c]\right)\sqrt{n_c}.
\end{equation}

When $V=0$, the following system for the super-fluid of the condensate can be obtained
\begin{equation}\label{BECSuperFluid}
\begin{aligned}
 {\partial_t n_c}\ +  \ \nabla_r\cdot(n_cv_c)\
=&\  - \ \Gamma_{12}[f]\\
\partial_t v_c +\frac{\nabla_r v_c^2}{2} = & \ -\nabla_r\mu_c.
\end{aligned}
\end{equation}

The potential $U$ and the condensate energy $\mathcal{E}_c$ are  written as follows
\begin{equation}\label{def-U}
U(t,r)\ = \ V(r) + 2g[n_c(t,r)+n_n(t,r)],
\end{equation}
and 
\begin{equation}\label{def-Ec}
\mathcal{E}_c(t,r) \ = \ \mu_c(t,r) \ +\ \frac{mv_c^2(t,r)}{2}.
\end{equation}
For the sake of simplicity, we suppose that $V\equiv 0$ and define the differential quantity
\begin{equation}\label{Diff}
\bar{d}p=\frac{dp}{(2\pi \hbar)^3}.
\end{equation} 
Notice that \eqref{C12} describes collisions between the condensate and the non-condensate atoms (condensate growth term) and \eqref{C22} describes collisions between non-condensate atoms. 
\begin{remark}
At temperature $T$, bosons  of mass $m$ can be regarded as quantum-mechanical
wavepackets which have an extent on the order of a thermal de Broglie wavelength $\lambda_{dB} =
\left(\frac{2 \pi \hbar^2}{m k_B T}\right)^\frac12$, where $k_B$ is the Boltzmann constant. The de Broglie wavelength $\lambda_{dB}$ describes the position uncertainty associated with the thermal momentum
distribution.  When the gas temperature is high $T>T_{BEC}$, $\lambda_{dB}$ is very small and  the weakly interacting gas can be treated
as a system of ``billiard balls'' (cf. \cite{durfee1998experimental,ketterle1999making}). The dynamics of the gas is described by the Boltzmann-Norheim (Uehling-Ulenbeck) equation, whose operator sometimes reads (cf. \cite{UehlingUhlenbeck:TPI:1933})
\begin{equation}\label{QBHightT}
\begin{aligned}
\mathcal{C}_{22}[f](t,r,p_1)\ =& \ \iiint_{\mathbb{R}^{3}\times\mathbb{R}^{3}\times\mathbb{R}^{3}}\delta({p}_1+{p}_2-{p}_3-{p}_4)\delta(\mathcal{E}_{{p}_1}+\mathcal{E}_{{p}_2}-\mathcal{E}_{{p}_3}-\mathcal{E}_{{p}_4})\times\\
\ & \ \times [(1+\vartheta f_1)(1+\vartheta f_2)f_3f_4-f_1f_2(1+\vartheta f_3)(1+ \vartheta f_4)]d p_2d p_3d p_4,
\end{aligned}
\end{equation}
where $\vartheta$ is proportional to $\hbar^3$. In the semiclassical limit, as $\vartheta$ tends to $0$, the quantum Boltzmann collision operator becomes the classical one. This means  at  high temperature, the behavior of the ``billiard balls''  Bose gas is, in some sense, still very similar to classical gases.

At the BEC transition temperature, $\lambda_{dB}$ becomes comparable to the distance between
atoms. As a result, the atomic wavepackets ``overlap'' and the indistinguishability of atoms becomes important. At this
temperature, bosons undergo a quantum-mechanical phase transition and the Bose-Einstein condensate is formed (cf. \cite{durfee1998experimental,ketterle1999making}). When the temperature of the gas is finite $T_{BEC}>T>0$K, the trapped Bose gas is composed of two distinct components: the high-density condensate, being localized at the center
of the trapping potential, and the low-density
cloud of thermally excited atoms, spreading over a much wider region. The dynamics of the thermal
cloud atoms is described by the kinetic equation \eqref{QBFull}. At this low temperature, the de Broglie wavelength of the excited atoms is very large, in comparison with the high temperature boson de Broglie wavelength. As a consequence, the thermal cloud kinetic equation cannot be treated 
as a system of ``billiard balls'' anymore. This explains the difference between the forms of the two collision operators $C_{22}$ and $\mathcal{C}_{22}$.

Note that, different from  classical Boltzmann collision operators, where the collision kernels are functions depending on the types of particles considered, the derived collision kernel for the quantum Boltzmann collision operator for bosons is 1 (cf. \cite{EscobedoVelazquez:2015:FTB}) when $T>T_{BEC}$. 
\end{remark}

\subsection{Scalings}

Different from the thesis \cite{Allemand:Thesis:2009}, in which the two collision operators $C_{12}$ and $C_{22}$ are assumed to have the same equilibrium distribution function, we follow \cite{GriffinNikuniZaremba:2009:BCG} to consider the most general regime, where excited atoms in the condensate need not to be in local  equilibrium with the condensate atoms. As a consequence, $C_{12}$ and $C_{22}$ in general do not share the same equilibrium distribution. A comparison between our results and the result of \cite{Allemand:Thesis:2009} will be discussed in details in Section \ref{Comparison}. Relying on these physical assumptions, we propose a new approach to obtain new Euler and Navier-Stokes approximations of the system.

It is known that the dynamics of the trapped Bose gases depends on its temperature $T$. Let us restrict our attention to the case where  $T$  is smaller but very close to the Bose-Einstein critical temperate $T_{BEC}$. At this temperature regime, the collisions between excited atoms are  rapid to establish a local equilibrium within the non-condensate component. As a consequence, the collision operator $C_{22}$ can be assumed to be  stronger than the collision operator $C_{12}$. This regime is often called {\it  the state of partial local equilibrium} which arises near $T_{BEC}$ when the density of the condensate is small.

Following \cite{GriffinNikuniZaremba:2009:BCG},  we  define the {\it static equilibrium} of the system
\begin{equation}\label{StaticEquilibrium}
\mathcal{F}_0(p) \ = \ \frac{1}{e^{\beta_0[(p-mv_{n0})^2/(2m) + U_0 - \mu_0]}-1},
\end{equation}
where $\beta_0$ is the static temperature parameter, $v_{n0}$ is the static fluid velocity, $\mu_0$ is the static chemical potential, $U_0$ is the static mean field. We also set the static density to be 
\begin{equation}\label{FluidDensity}
 {n_n}_0 \ =  \ \int_{\mathbb{R}^3}\mathcal{F}_0(p) \bar{d} p.
\end{equation}

Note that when $T$ is sufficiently close to $T_{BEC}$, the bosons  are in the  {\it particle-like regime}, i.e. they behave like particles. Let us also mention that when temperature $T$ is very close to $0$, the bosons will be in the phonon-like regime (cf. \cite{ReichlBook}). Since we are interested in the behavior of the particles when $T$ is close to $T_{BEC}$, let us define the collision frequency with respect to $C_{12}$
\begin{equation}\label{CollisionFrequencyC12}
\begin{aligned}
\nu_{12}(p_1) \ =  \ &  \frac{\lambda_1 n_c}{m^2} 
\iint_{\mathbb{R}^3\times\mathbb{R}^3}\delta(mv_c + p_1 -p_2-p_3)\delta(\mathcal{E}_c+\mathcal{E}_{p_1}-\mathcal{E}_{p_2}-\mathcal{E}_{p_3})\times\\
&\times[\mathcal{F}_0(p_2)+\mathcal{F}_0(p_3)+1]dp_2dp_3\ + \ \\
& \ +2\frac{\lambda_1 n_c}{m^2} 
\iint_{\mathbb{R}^3\times\mathbb{R}^3}\delta(mv_c + p_2 -p_1-p_3)\delta(\mathcal{E}_c+\mathcal{E}_{p_2}-\mathcal{E}_{p_1}-\mathcal{E}_{p_3})\mathcal{F}_0(p_3)dp_2dp_3,
\end{aligned}
\end{equation}
as well as the associated mean collision frequency:
\begin{equation}\label{MeanCollisionFrequencyC12}
\bar{\nu}_{12}\ = \ \frac{1}{{n_{n_0}}m}\int_{\mathbb{R}^3}\nu_{12}(p)\mathcal{F}_0(p)\bar{d}p.
\end{equation}
The inverse of $\nu_{12}(p) $ and $\bar{\nu}_{12}$ are defined to be, respectively, the free time $\tau_{12}(p)$ and the mean field time $\bar\tau_{12}$: 
\begin{equation}\label{MeanFieldTimeC12}
\tau_{12}(p) \  = \ \frac{1}{\nu_{12}(p)}, \ \ \ \ \ \bar\tau_{12} \ = \ \frac{1}{\bar{\nu}_{12}}.
\end{equation}
We now determine the average speed of the particles 
\begin{equation}\label{AvergeSpeed}
\bar{c} \ = \ \frac{1}{{n_n}_0m}\int_{\mathbb{R}^3}\sqrt{p^2}\mathcal{F}_0(p)\bar{d}p,
\end{equation}
and the mean free path 
\begin{equation}\label{MeanFreePathC12}
l_{12} \ = \ \bar{c}\bar{\tau}_{12}.
\end{equation}

Similarly, the collision frequency and the mean collision frequency associated to $C_{22}$ can be defined 
 \begin{equation}\label{CollisionFrequencyC22}
\begin{aligned}
\nu_{22}(p_1) \ =  \ &  \frac{\lambda_2}{{n_n}_0m} 
\iint_{\mathbb{R}^3\times\mathbb{R}^3\times\mathbb{R}^3}\delta(p_1 + p_2-p_3-p_4)\delta(\mathcal{E}_{p_1}+\mathcal{E}_{p_2}-\mathcal{E}_{p_3}-\mathcal{E}_{p_4})\times\\
& \times \mathcal{F}_0(p_2)(1+\mathcal{F}_0(p_3))(1+\mathcal{F}_0(p_4))dp_2dp_3dp_4,
\end{aligned}
\end{equation}
and
\begin{equation}\label{MeanCollisionFrequencyC12}
\bar{\nu}_{22}\ = \ \frac{1}{{n_n}_0m}\int_{\mathbb{R}^3}\nu_{22}(p)\mathcal{F}_0(p)\bar{d}p.
\end{equation}
We also define the free time $\tau_{22}(p)$, the mean field time $\bar\tau_{22}$ and the mean free path $l_{22}$
\begin{equation}\label{MeanFieldTimeC12}
\tau_{22}(p) \  = \ \frac{1}{\nu_{22}(p)}, \ \ \ \ \ \bar\tau_{22} \ = \ \frac{1}{\bar{\nu}_{22}}, \ \ \ \ \ \ l_{22} \ = \ \bar{c}\bar{\tau}_{22}.
\end{equation}

Let $L$ and $\theta$ be the reference length and time, respectively. Following \cite{Sone:KTN:2002,BouchutGolse:2000:KEA}, we introduce the rescaled variables 
\begin{equation}\label{RescaledVariables}
\tilde{r}=\frac{r}{L}, \ \ \ \tilde{t}=\frac{t}{\theta}, \ \ \ \tilde{p}=\frac{p}{P}, P=m\bar{c}, \ \ \ \tilde{v}_c=\frac{v_c}{\bar{c}}. 
\end{equation}

Note that under this scaling, 
\begin{equation}\label{nnnewscaling}
 n_n(t,r) \ =  \ \int_{\mathbb{R}^3}f(t,r,p)d p \ =  \ P^3\int_{\mathbb{R}^3}f(t,r,\tilde{p})d \tilde{p}.
\end{equation}

We also rescale $U$ as $\tilde{U}=U/U_0$, where $U_0$ is the reference potential field. Define 
\begin{eqnarray}\nonumber
\tilde{C}_{12}[f](t,r,\tilde{p}_1)&:=&\tilde{\lambda}_1n_c(t,r) \iint_{\mathbb{R}^{3}\times\mathbb{R}^{3}}\delta(\tilde{v}_c+ \tilde{p}_1-\tilde{p}_2-\tilde{p}_3)\delta(\mathcal{E}_c+\mathcal{E}_{\tilde{p}_1}-\mathcal{E}_{\tilde{p}_2}-\mathcal{E}_{\tilde{p}_3})\\\label{RescaledC12}
& &\times[(1+f_1)f_2f_3-f_1(1+f_2)(1+f_3)]d \tilde{p}_2d \tilde{p}_3\\\nonumber
&&-2\tilde{\lambda}_1n_c(t,r)\iint_{\mathbb{R}^{3}\times\mathbb{R}^{3}}\delta(\tilde{v}_c+\tilde{p}_2-\tilde{p}_1-\tilde{p}_3)\delta(\mathcal{E}_c+\mathcal{E}_{\tilde{p}_2}-\mathcal{E}_{\tilde{p}_1}-\mathcal{E}_{\tilde{p}_3})\\\nonumber
& &\times[(1+f_2)f_1f_3-f_2(1+f_1)(1+f_3)]d \tilde{p}_2d \tilde{p}_3,\\\label{RescaledC22}
\tilde{C}_{22}[f](t,r,\tilde{p}_1)&:=&\tilde{\lambda}_2\iiint_{\mathbb{R}^{3}\times\mathbb{R}^{3}\times\mathbb{R}^{3}}\delta(\tilde{p}_1+\tilde{p}_2-\tilde{p}_3-\tilde{p}_4)\\\nonumber
& &\times\delta(\mathcal{E}_{\tilde{p}_1}+\mathcal{E}_{\tilde{p}_2}-\mathcal{E}_{\tilde{p}_3}-\mathcal{E}_{\tilde{p}_4})\times\\\nonumber
&&\times [(1+f_1)(1+f_2)f_3f_4-f_1f_2(1+f_3)(1+f_4)]d \tilde{p}_2d \tilde{p}_3d \tilde{p}_4,
\end{eqnarray} where
\begin{equation}\label{LambdaC12}
 \tilde{\lambda}_{1}=P^2\lambda_{1}/\bar{c},
\end{equation}
and
\begin{equation}\label{LambdaC22}
\tilde{\lambda}_{2}=P^5\lambda_{2}/\bar{c}.\end{equation}
As a consequence, we can define the rescaled mean free paths and the rescaled mean field times to be
$$\tilde{l}_{22}=\frac{l_{22}}{P^5}, \ \  \tilde{\tau}_{22}=\frac{\bar{\tau}_{22}}{P^5},$$
and
$$\tilde{l}_{12}=\frac{l_{12}}{P^2}, \ \  \tilde{\tau}_{12}=\frac{\bar{\tau}_{12}}{P^2}.$$
We also set
\begin{equation}\label{Chat}
\hat{C}_{12}[f]: =\tilde{l}_{12} \tilde{C}_{12}[f], \ \ \ \hat{C}_{22}[f]: =\tilde{l}_{22}\tilde{C}_{22}[f].
\end{equation}
The following rescaled version of \eqref{QBFull} then follows:
\begin{equation}\label{RescaledQB1}
\frac{\sqrt{\tilde{l}_{12}\tilde{l}_{22}}}{\theta\bar{c}}  \partial_{\tilde{t}}f \ + \ \frac{\sqrt{\tilde{l}_{12}\tilde{l}_{22}}}{L}\frac{P}{m\bar{c}}\tilde{p}\cdot \nabla_{\tilde{r}}f \ - \ \frac{\sqrt{\tilde{l}_{12}\tilde{l}_{22}}}{L}\frac{U_0}{P\bar{c}}\nabla_{\tilde{r}}\tilde{U}\cdot \nabla_{\tilde{p}} f \ = \ \sqrt{\frac{\tilde{l}_{22}}{\tilde{l}_{12}}}\hat{C}_{12}[f] \ + \ \sqrt{\frac{\tilde{l}_{12}}{\tilde{l}_{22}}}\hat{C}_{22}[f].
\end{equation}

Notice that $\frac{\tilde\tau_{22}}{\tilde\tau_{12}}=\frac{\tilde{l}_{22}}{\tilde{l}_{12}}$ is a dimensionless parameter and is proportional to $\frac{\tilde\lambda_1}{\tilde\lambda_2}$.

In this paper, we will consider two hydrodynamic approximations: Euler and Navier-Stokes. 
\begin{itemize}
\item The Euler approximation is quite general and valid under  a general physical situation. The collisions between excited atoms are  fast to establish a local equilibrium within the non-condensate component, and the quantity $\tilde\tau_{22}$ is smaller than $\tilde\tau_{12}$ but the ratio between $ \tilde\tau_{22}$ and $\tilde\tau_{12}$ is not necessarily very small. 
\item The Navier-Stokes approximation is valid under the physical assumption that the collisions between excited atoms are extremely rapid to establish a local equilibrium within the non-condensate component and   $\tilde\tau_{22}<<\tilde\tau_{12}$. 
\end{itemize}
We suppose $\frac{\tilde\tau_{22}}{\tilde\tau_{12}}=\epsilon^{2}$. {\it The Euler approximation is valid in any physical assumption and we do not need to impose the assumption that $\epsilon$ is small, then $\epsilon$ is just a parameter. In the Navier-Stokes approximation, we need to impose the assumption that $\epsilon$ is small and then we will use it as the small parameter in the usual Chapman-Enskog expansion process. }

The constants $\frac{\sqrt{\tilde{l}_{12}\tilde{l}_{22}}}{\theta\bar{c}}$,  $\frac{\sqrt{\tilde{l}_{12}\tilde{l}_{22}}}{L}$ can be set to be $1$ by rescaling again the space and time variables $\tilde{t}\to \frac{\sqrt{\tilde{l}_{12}\tilde{l}_{22}}}{\theta\bar{c}}\tilde{t}$, $\tilde{r}\to \frac{\sqrt{\tilde{l}_{12}\tilde{l}_{22}}}{L}\tilde{r}$, and note that $\frac{P}{m\bar{c}}=1$, we obtain the following equation
\begin{equation}\label{RescaledQB2}
\partial_{\tilde{t}}f \ + \ \tilde{p}\cdot \nabla_{\tilde{r}}f \ - \ \frac{U_0}{m\bar{c}^2}\nabla_{\tilde{r}}\tilde{U}\cdot \nabla_{\tilde{p}} f \ = \ \epsilon\hat{C}_{12}[f] \ + \ \frac{1}{\epsilon}\hat{C}_{22}[f].
\end{equation}

Notice that $g$ is also the principle small parameter used in the derivation of the system \eqref{GP}-\eqref{QBFull}. Indeed, the derivation starts with the usual Heisenberg equation of motion for the quantum field
operator. The equation for the condensate wavefunction follows by averaging the Heisenberg equation with respect to a broken-symmetry nonequilibrium ensemble. Taking the difference between the Heisenberg equation and the  equation for the condensate wavefunction and keeping only the terms of low orders with respect to $g$, we obtain the equation of the noncondensate field operator, which, by a Wigner transform, leads to the quantum Boltzmann equation. In this process, one computes the collision integrals $C_{12}$, $C_{22}$ to second order $O(g^2)$ in $g$  and keep interaction effects in the excitation energies and chemical potential only to first order $O(g)$. For  a more detailed explanation of this procedure, we refer to, for instance, Sections 3.1, 3.2 and  5.3 of the book \cite{GriffinNikuniZaremba:2009:BCG}. {\it Since $U_0$ has to be chosen  proportional to $g$,  the dimensionless parameter $\frac{U_0}{m\bar{c}^2}$ might be considered to be small and set it to be $\tilde{g}=\epsilon^{\delta_0}$, $0<\delta_0<1$ in Section \ref{Sec:NavierStokes}, where the Chapman-Enskog expansion is used.} 

The equation then follows, as a result of the previous scaling
\begin{equation}\label{RescaledQBFinal}
\partial_{\tilde{t}}f \ + \ \tilde{p}\cdot \nabla_{\tilde{r}}f \ - \ \tilde{g}\nabla_{\tilde{r}}\tilde{U}\cdot \nabla_{\tilde{p}} f \ = \ \ \epsilon\hat{C}_{12}[f] \ + \ \frac{1}{\epsilon}\hat{C}_{22}[f].
\end{equation}

Under this scaling, the Gross-Pitaevski equation also becomes
\begin{equation}
\begin{aligned}\label{RescaledQB4}
i \frac{\hbar}{\theta} {\partial_{\tilde{t}} {\Phi}(t,r)} =&\ \Big(-\frac{\hbar^{2} \Delta_{\tilde{r}}}{2m L^2}+g[n_c(t,r) + 2{n_n}(t,r)] -\frac{i}{\tilde\tau_{12}}\tilde{\Lambda}_{12}[f](t,r)\Big){\Phi}(t,r),
\end{aligned}
\end{equation}
where $$\tilde{\Lambda}_{12}[f]=\frac{\hbar}{2n_c}\int_{\mathbb{R}^3}\hat{C}_{12}[f]\bar{d}p.$$
By the same argument as above, we also obtain
\begin{equation}
\begin{aligned}\label{RescaledQB4b}
i \frac{\sqrt{\tilde{l}_{12}\tilde{l}_{22}}}{\theta\bar{c}} {\partial_{\tilde{t}} {\Phi}(t,r)} =&\ \Big(-\frac{\hbar}{mL}\frac{\sqrt{\tilde{l}_{12}\tilde{l}_{22}}\Delta_{\tilde{r}}}{2 L\bar{c}}+\frac{\sqrt{\tilde{l}_{12}\tilde{l}_{22}}U_0}{\hbar\bar{c}}\tilde{U}_*(t,r)-\frac{i \sqrt{\tilde{l}_{12}\tilde{l}_{22}}}{\tilde{l}_{12}}{\tilde{\Lambda}_{12}[f](t,r)}\Big){\Phi}(t,r),
\end{aligned}
\end{equation}
where $\tilde{U}_*=U_*/U_0$ and  $U_*(t,r)=g[n_c(t,r) + 2{n_n}(t,r)]$, since $U_*(t,r)$ has the dimension of $U(t,r)$.

Notice that $\frac{\hbar}{m\bar{c}}$ has the dimensions of a length (Compton wavelength) and $\frac{\hbar}{mL\bar{c}}$ is dimensionless; hence the quantity $\frac{\hbar}{mL}\frac{\sqrt{\tilde{l}_{12}\tilde{l}_{22}}}{2 L\bar{c}}$ is dimensionless.  Moreover, $\frac{\sqrt{\tilde{l}_{12}\tilde{l}_{22}}U_0}{\hbar\bar{c}}$ is the product of the three dimensionless parameters $\frac{\sqrt{\tilde{l}_{12}\tilde{l}_{22}}}{L}$, $\frac{mL\bar{c}}{\hbar}$ and $\frac{U_0}{m\bar{c}^2}=\tilde{g}$. Setting all of the dimensionless parameter to be $1$ by the same rescaling argument used for \eqref{RescaledQB2} and dropping the tilde and hat signs
\begin{equation}
\begin{aligned}\label{RescaledGPFinal}
i {\partial_{{t}} {\Phi}} =&\ \Big(-\frac{\Delta_{{r}}}{2}+{g}U_* -i\epsilon{\Lambda}_{12}[f]\Big){\Phi},
\end{aligned}
\end{equation}
where $g$ stands for the dimensionless parameter $\tilde{g}=\epsilon^{\delta_0}$. When $V=0$, the following system for the super-fluid of the condensate can be deduced
\begin{equation}\label{RescaledSystem2b}
\begin{aligned}
 {\partial_t n_c}\ +  \ \nabla_r\cdot(n_cv_c)\
=&\  -  \epsilon \Gamma_{12}[f]\\
\partial_t v_c +\frac{\nabla_r v_c^2}{2} = & \ -\nabla_r\mu_c.
\end{aligned}
\end{equation}
 we then obtain
 the system
\begin{equation}
\begin{aligned}\label{RescaledSystem}
\partial_{{t}}f \ + \ {p}\cdot \nabla_{{r}}f \ - & \ g\nabla_{{r}}{U}\cdot \nabla_{{p}} f \ =  \ \ \epsilon {C}_{12}[f] \ + \ \frac{1}{\epsilon}{C}_{22}[f], (0<\delta_0<1), \\
i {\partial_t {\Phi}} =&\ \Big(-\frac{\Delta_{{r}}}{2}+g[n_c + 2{n_n}] -i\epsilon\Lambda_{12}[g]\Big){\Phi}.
\end{aligned}
\end{equation}
We recall below the formulas for $C_{12}$, $C_{22}$ and $\Lambda_{12}$
\begin{eqnarray}\nonumber
{C}_{12}[f](t,r,{p}_1)&=&n_c(t,r) \iint_{\mathbb{R}^{3}\times\mathbb{R}^{3}}\delta(v_c+ {p}_1-{p}_2-{p}_3)\delta(\mathcal{E}_c+\mathcal{E}_{{p}_1}-\mathcal{E}_{{p}_2}-\mathcal{E}_{{p}_3})\\\label{C12Dimless}
& &\times[(1+f_1)f_2f_3-f_1(1+f_2)(1+f_3)]d {p}_2d {p}_3\\\nonumber
&&-2n_c(t,r)\iint_{\mathbb{R}^{3}\times\mathbb{R}^{3}}\delta(v_c+{p}_2-{p}_1-{p}_3)\delta(\mathcal{E}_c+\mathcal{E}_{{p}_2}-\mathcal{E}_{{p}_1}-\mathcal{E}_{{p}_3})\\\nonumber
& &\times[(1+f_2)f_1f_3-f_2(1+f_1)(1+f_3)]d {p}_2d {p}_3,\\\label{C22Dimless}
{C}_{22}[f](t,r,{p}_1)&=&\iiint_{\mathbb{R}^{3}\times\mathbb{R}^{3}\times\mathbb{R}^{3}}\delta({p}_1+{p}_2-{p}_3-{p}_4)\delta(\mathcal{E}_{{p}_1}+\mathcal{E}_{{p}_2}-\mathcal{E}_{{p}_3}-\mathcal{E}_{{p}_4})\times\\\nonumber
&&\times [(1+f_1)(1+f_2)f_3f_4-f_1f_2(1+f_3)(1+f_4)]d {p}_2d {p}_3d {p}_4,\\
\Lambda_{12}[f](t,r) & = &\ \frac{1}{n_c(t,r)}\int_{\mathbb{R}^{3}}C_{12}[f](t,r,p){d p}.
\end{eqnarray}
We also define the  differential operators
\begin{equation}\label{Df}
\mathcal{D}f \ = \ {\partial_t f}\ +\ {p}\cdot\nabla_{{r}} f \ - \ g\nabla_r U\cdot \nabla_p f\ - \epsilon C_{12}[f],
\end{equation}
\begin{equation}\label{Df2}
\mathbb{D}f \ = \ {\partial_t f}\ +\ {p}\cdot\nabla_{{r}} f \ - \ g\nabla_r U\cdot \nabla_p f,
\end{equation}
\begin{equation}\label{Df3}
\Pi f \ = \ {\partial_t f}\ +\ {p}\cdot\nabla_{{r}} f,
\end{equation}
and then get 
\begin{equation}
\begin{aligned}\label{RescaledSystem2a}
\mathbb{D}f \ =  &  \ \ \epsilon {C}_{12}[f] \ + \ \frac{1}{\epsilon}{C}_{22}[f], (0<\delta_0<1).
\end{aligned}
\end{equation}

The new constant $\epsilon$ is the small parameter that we will use in the usual Chapman-Enskog expansion process  in Section \ref{Sec:NavierStokes}.

\section{Properties of the collision operators}\label{Sec:Operators}
In this section, we study the main properties of the two collision operators $C_{12}$ and $C_{22}$.
\subsection{Collision invariants and equilibrium of $C_{22}$}
Let us start with $C_{22}$, which can be represented as:
\begin{equation}
\label{C22Presentation}
C_{22}[f] \ = \ B_1[f,f]  \ + \ B_2[f,f,f],
\end{equation}
in which
\begin{equation}
\label{Q1}
\begin{aligned}
B_1[f,g] \ = & \ \frac{1}{2}\iiint_{\mathbb{R}^{3}\times\mathbb{R}^{3}\times\mathbb{R}^{3}}\delta({p}_1+{p}_2-{p}_3-{p}_4)\delta(\mathcal{E}_{{p}_1}+\mathcal{E}_{{p}_2}-\mathcal{E}_{{p}_3}-\mathcal{E}_{{p}_4})\times\\
&\times [f_3g_4+f_4g_3-f_1g_2-f_2g_1]d p_2d p_3d p_4,
\end{aligned}
\end{equation}
and
\begin{equation}
\label{Q2}
\begin{aligned}
B_2[f,g,h] \ = & \ \frac{1}{6}\iiint_{\mathbb{R}^{3}\times\mathbb{R}^{3}\times\mathbb{R}^{3}}\delta({p}_1+{p}_2-{p}_3-{p}_4)\delta(\mathcal{E}_{{p}_1}+\mathcal{E}_{{p}_2}-\mathcal{E}_{{p}_3}-\mathcal{E}_{{p}_4})\times\\
&\times [f_3g_4h_1+f_4g_3h_1+f_3g_4h_2+f_4g_3h_2\\
&\ \ +f_1g_4h_3+f_1g_3h_4+f_2g_4h_3+f_3g_3h_4\\
&\ \ +f_4g_1h_3+f_3g_1h_4+f_4g_2h_3+f_3g_2h_4\\
&\ \ -f_1g_2h_3-f_2g_1h_3-f_1g_2h_4-f_2g_1h_4\\
&\ \ -f_3g_1h_2-f_3g_2h_1-f_4g_1h_2-f_4g_2h_1\\
&\ \ -f_1g_3h_2-f_2g_3h_1-f_1g_4h_2-f_2g_4h_1]d p_2d p_3d p_4,
\end{aligned}
\end{equation}
where we have used the same notations $f_1$, $f_2$, $f_3$, $f_4$, $g_1$, $g_2$ , $g_3$, $g_4$, $h_1$, $h_2$ , $h_3$, $h_4$ with the ones used in \eqref{QBFull}.

The operator $C_{22}$ shares some important features with the classical Boltzmann collision operator. Among these features, the following can be proved by switching the variables $(p_1,p_2)\leftrightarrow (p_2,p_1)$,  $(p_1,p_2)\leftrightarrow (p_3,p_4)$, in the integrals of  $B_1$ as in the classical case (cf. \cite{Villani:2002:RMT}):
\begin{equation}\label{Q1Conservation}
\int_{\mathbb{R}^3}\Psi_i(p)B_1[f,g](p)d p \ = \ 0, \ \ \ i=0,1,2,3,4,
\end{equation}
where 
\begin{equation}\label{NullSpace}
\Psi_0(p) \ = \  1, \ \ \Psi_i(p) \ = \ p^i, \ \ (i=1,2,3), \ \ \Psi_4(p) \ = \ |p|^2,
\end{equation}
are the collision invariants and $p^i$ is the $i$-th component of the vector $p=(p^1,p^2,p^3)$.

Moreover, we also have
\begin{equation}\label{Q1Conservation}
\int_{\mathbb{R}^3}\Psi_i(p)B_2[f,g,h](p)d p \ = \ 0, \ \ \ i=0,1,2,3,4.
\end{equation}

Similar as the classical Boltzmann collision operator, $C_{22}$ also has a local equilibrium of the form
\begin{equation}\label{Sec:EulerLimit:E27}
\mathcal{F}(t,r,p) \ = \ \frac{1}{e^{\beta[(p-v_n)^2/2 + U - \mu]}-1},
\end{equation}
where $\beta(t,r)$ is the temperature parameter, $v_n(t,r)$ is the local fluid velocity, $\mu(t,r)$ is the local chemical potential (which is different from the condensate chemical potential $\mu_c(t,r)$ defined in \eqref{def-muc}), $U(t,r)$ is the mean field. Then
$$C_{22}[\mathcal{F}] \ = \ 0.$$
Let us now define the following Gaussian 
\begin{equation}\label{Gaussian}
\mathcal{M}(t,r,p) \ = \ \gamma(t,r)e^{-\frac{|p-u(t,r)|^2}{2\tau(t,r)}},
\end{equation}
where 
\begin{equation}
\gamma(t,r) \ = \   e^{\beta(U(t,r)-\mu(t,r))},\ \ \
u(t,r) \ = \  v_n(t,r),\ \ \
\tau(t,r)\  = \  \frac{1}{\beta(t,r)}. 
\end{equation}
The local equilibrium $\mathcal{F}$ can be expressed in terms of $\mathcal{M}$ as
\begin{equation}\label{Maxwellian}
\mathcal{F}(t,r,p) \ = \ \frac{\mathcal{M}(t,r,p)}{1-\mathcal{M}(t,r,p)}.
\end{equation}

Note that $u$ is a vector $u=(u_1,u_2,u_3)$.

\subsection{Linearized operator of $C_{22}$}

Let $L^2(\mathbb{R}^3)$ be the space of real, measurable functions, whose second power is integrable on $\mathbb{R}^3$, with the norm $\|\cdot\|_{L^2}$ and inner product $(,)_{L^2}$. We consider the linearized operator of $C_{22}$ around a fixed equilibrium $\mathcal{F}(t,r,p)$, which, by a classical process can be  defined as
\begin{equation}\label{C22Linear1}
\mathcal{L} \ : = \ 2B_1(\mathcal{F},\cdot) \ + \ 3B_2(\mathcal{F},\mathcal{F},\cdot),
\end{equation}
or equivalently
\begin{equation}\label{C22Linear2}
\begin{aligned}
\mathcal{L}(\mathcal{F}f)(t,r,p_1) \ = & \ \int_{\mathbb{R}^3\times\mathbb{R}^3\times\mathbb{R}^3}\delta(p_1+p_2-p_3-p_4)\delta(\mathcal{E}_{p_1}+\mathcal{E}_{p_2}-\mathcal{E}_{p_3}-\mathcal{E}_{p_4})\\
\ & \ \times \frac{\mathcal{M}_1\mathcal{M}_2}{(1-\mathcal{M}_1)(1-\mathcal{M}_2)(1-\mathcal{M}_3)(1-\mathcal{M}_4)}\times\\
\ &\ \times \Big[(1-\mathcal{M}_3)f(p_3)+(1-\mathcal{M}_4)f(p_4) - (1-\mathcal{M}_2)f(p_2)\\
\ & \ - (1-\mathcal{M}_1)f(p_1)\Big]d p_2d p_3d p_4,
\end{aligned}
\end{equation}
for some function $f(p)$ and fixed values $(t,r)\in\mathbb{R}_+\times\mathbb{R}^3$ and we employ the shorthand notations $\mathcal{M}_i=\mathcal{M}(t,r,p_i)$, $i=1,2,3,4$. Now, let us consider the inner product between the above linearized operator and some test function $\varphi$. The classical argument (cf. \cite{Villani:2002:RMT}) for the classical linearized Boltzmann collision operator can be applied and gives:
\begin{equation*}
\begin{aligned}
\left(\frac{\mathcal{M}}{\mathcal{F}}\varphi,\mathcal{L}(\mathcal{F}f)\right)_{L^2} \ =& \ -\frac{1}{4}\int_{\mathbb{R}^3\times\mathbb{R}^3\times\mathbb{R}^3\times\mathbb{R}^3}\delta(p_1+p_2-p_3-p_4)\delta(\mathcal{E}_{p_1}+\mathcal{E}_{p_2}-\mathcal{E}_{p_3}-\mathcal{E}_{p_4}) \\
\ & \ \times \frac{\mathcal{M}_1\mathcal{M}_2}{(1-\mathcal{M}_1)(1-\mathcal{M}_2)(1-\mathcal{M}_3)(1-\mathcal{M}_4)}\times\\
\ &\ \times \Big[(1-\mathcal{M}_3)f(p_3)+(1-\mathcal{M}_4)f(p_4) - (1-\mathcal{M}_2)f(p_2)\\
\ & \ - (1-\mathcal{M}_1)f(p_1)\Big]\Big[(1-\mathcal{M}_3)\varphi(p_3)+(1-\mathcal{M}_4)\varphi(p_4)\\
\ &\ - (1-\mathcal{M}_2)\varphi(p_2) - (1-\mathcal{M}_1)\varphi(p_1)\Big]dp_1d p_2d p_3d p_4,\end{aligned}
\end{equation*} 
which implies
\begin{equation}\label{C22Linearizedpositivity}
\left(\frac{\mathcal{M}}{\mathcal{F}}f,\mathcal{L}(\mathcal{F}f)\right)_{L^2} \ \le \ 0,
\end{equation}
and
\begin{equation*}
\left(\frac{\mathcal{M}}{\mathcal{F}}\varphi,\mathcal{L}(\mathcal{F}f)\right)_{L^2} \ = \ \left(\frac{\mathcal{M}}{\mathcal{F}}f,\mathcal{L}(\mathcal{F}\varphi)\right)_{L^2},
\end{equation*}
for all function $\varphi$ and $f$ such that the integrals are well-defined. The equality in \eqref{C22Linearizedpositivity} holds true if and only if $\frac{\mathcal{M}}{\mathcal{F}}f$ is identical to one of the five functions defined in \eqref{NullSpace}. 

From the above observation, we are now able to define the kernel of the linearized collision operator $\mathcal{L}$ of $C_{22}$:
$$\mathcal{N} \ := \ \mathrm{ker}\mathcal{L} \ = \ \mathrm{span}\left\{\frac{\mathcal{F}^2}{\mathcal{M}}\Psi_i: \ i=0,\cdots,4\right\},$$
and its orthogonal space:
$$\mathcal{R} \ := \ \mathcal{N}^{\bot} \ = \ \left\{G\in L^2(\mathbb{R}^3)\ : \ \left(G,\frac{\mathcal{F}^2}{\mathcal{M}}\Psi_i\right)_{L^2}=0, \ i=0,\cdots,4\right\}.$$

On $L^2(\mathbb{R}^3)$, we also define the orthogonal projection operators $\mathbb{P}$ and $\mathbb{P}^{\bot}=1-\mathbb{P}$ on to $\mathcal{N}$ and $\mathcal{R}$. By normalizing $\{\Psi_i\}_{i=0,\cdots,4}$, we obtain the following orthonormal basis of the space $\mathcal{N}$
\begin{equation}\label{OrthonormalBasis}
\left\{\frac{\psi_i}{\sqrt{\omega_i}}\frac{\mathcal{F}^2}{\mathcal{M}}: \ \ \ i=0,\cdots,4\right\},
\end{equation}
with
$$\psi_0=1; \ \ \psi_{i}=p^i-u_i, \ \ i=1,2,3; \ \ \psi_4=|p-u|^2-6\tau\frac{\Omega_1(\gamma)}{\Omega_0(\gamma)},$$
\begin{equation*}
\begin{aligned}
\omega_0 \ = & \ \int_{\mathbb{R}^3}\frac{\mathcal{F}^2}{\mathcal{M}}d p \ = \ 2^{3/2}\pi\tau^{3/2}\gamma\Omega_0(\gamma);\\
\omega_i \ = & \ \int_{\mathbb{R}^3}\frac{\mathcal{F}^2}{\mathcal{M}}|\psi_i(p)|^2d p \ = \ 2^{5/2}\pi\tau^{5/2}\gamma\Omega_1(\gamma),\ \ i=1,2,3; \\
\omega_4 \ = & \ \int_{\mathbb{R}^3}\frac{\mathcal{F}^2}{\mathcal{M}}|\psi_4(p)|^2d p \ = \ 2^{7/2}\pi\tau^{7/2}\gamma\Sigma(\Omega_0(\gamma),\Omega_1(\gamma),\Omega_2(\gamma));
\end{aligned}
\end{equation*}
where
$$\Sigma(x,y,z) \ = \ \frac{5xz-9y^2}{x},$$
and
\begin{equation}\label{sigmafunction}
\Omega_k(\gamma) \ = \ \int_{0}^\infty \frac{y^{k-1/2}}{e^y+\gamma} dy, \ \ \ \ k>-1/2.
\end{equation}

\subsection{Hydrodynamics quantities}

In order to study the hydrodynamics limit of the system, let us define the following moments of the function $f(t,r,p)$:
\begin{equation}\label{FluidDensity}
 {n_n}[f](t,r) \ =  \ \int_{\mathbb{R}^3}f(t,r,p){d} p,
\end{equation}
\begin{equation}\label{FluildVelocity}
u[f](t,r)\ = v_n(t,r)[f](t,r)\ = \ \frac{1}{{n_n}[f](t,r)}\int_{\mathbb{R}^3} {p}f(t,r,p){d} p,
\end{equation}
\begin{equation}\label{FluidEnergy}
{\mathbb{E}_n}[f](t,r) \ = \ \frac{1}{2}\int_{\mathbb{R}^3}f(t,r,p)|p-v_n[f](t,r)|^2 {d} p,
\end{equation}
\begin{equation}\label{FluidEnergye}
{\tilde{\mathbb{E}}_n}[f](t,r)\ = \ \frac{2{\mathbb{E}_n}[f](t,r)}{3}, \ \ \  e_n[f](t,r) \ = \ \frac{{\tilde{\mathbb{E}}_n}[f](t,r)}{ {n_n}[f](t,r)}.
\end{equation}

Replacing $f$ by $\mathcal{F}$, we obtain
\begin{equation}\label{FluidDensityM}
\begin{aligned}
 {n_n}[\mathcal{F}] \ = &  \ 2^{5/2}\pi\tau^{3/2}\Omega_1(\gamma),\\
{\mathbb{E}_n}[\mathcal{F}] \ =  & \ {2^{5/2}}\pi \tau^{5/2}\gamma\Omega_2(\gamma),
\end{aligned}
\end{equation} 
where $\Omega_1,$ $\Omega_2$ are defined in \eqref{sigmafunction}.
For the sake of simplicity, we denote ${n_n}[\mathcal{F}], v_n[\mathcal{F}],$ $ u[\mathcal{F}], $ ${\mathbb{E}_n}[\mathcal{F}],$ $\tilde{\mathbb{E}}_n[\mathcal{F}]$, $ e_n[\mathcal{F}]$ by ${n_n}, v_n, u, {\mathbb{E}_n}, \tilde{\mathbb{E}}_n$, and $e_n$.

We indeed can compute $\gamma$ and $\tau$ as
\begin{equation}
\label{gamma}
\gamma \ = \ \left(\frac{\mathrm{Id}\Omega_2}{\Omega_1^{5/3}}\right)^{-1}\left(\frac{2^{5/3}\pi^{2/3}{\mathbb{E}_n}}{{n_n}^{5/3}}\right),
\end{equation}
and
\begin{equation}
\label{tau}
\tau \ = \ \left(\frac{{n_n}}{2^{5/2}\pi\Omega_1\left(\left(\frac{\mathrm{Id}\Omega_2}{\Omega_1^{5/3}}\right)^{-1}\left(\frac{2^{5/3}\pi^{2/3}{\mathbb{E}_n}}{{n_n}^{5/3}}\right)\right)}\right)^{2/3}.
\end{equation}

\subsection{Computing $\Gamma_{12}[\mathcal{F}]$}
Now, let us consider the collision operator $C_{12}$. This operator also has the collision invariant property:
\begin{equation}\label{C12Conservation}
\int_{\mathbb{R}^3}(\Psi_i(p) - v_{ci}) C_{12}[f]d p \ = \ \int_{\mathbb{R}^3}\left(\Psi_4(p)+ {2}U- {2}\mu_c-{v_c^2}\right)C_{12}[f]d p\ = \ 0, \ \ \ i=1,2,3.
\end{equation}

An important property of $C_{12}$ is that $\mathcal{F}$ is not an equilibrium of $C_{12}$. We have:
\begin{equation}\label{Gamma12a}
\begin{aligned}
&\Gamma_{12}[\mathcal{F}]\ := \ \int_{\mathbb{R}^3}C_{12}[\mathcal{F}]{d}p \ \ \\  
& = \ -n_c[1-e^{-\beta(\mu-\mu_c -(v_n-v_c)^2/2)}]\iiint_{\mathbb{R}^3\times\mathbb{R}^3\times\mathbb{R}^3}\delta(v_c+p_1-p_2-p_3)\times\\
& \ \times \delta(\mathcal{E}_c+\mathcal{E}_{p_1}-\mathcal{E}_{p_2}-\mathcal{E}_{p_3})(1+\mathcal{F}(t,r,p_1))\mathcal{F}(t,r,p_2)\mathcal{F}(t,r,p_3)dp_1dp_2dp_3.
\end{aligned}
\end{equation}
Expanding $\mathcal{F}$ into Taylor series of $\mathcal{M}$, we can simplify the above integral as
\begin{equation}\label{Gamma12b}
\begin{aligned}
&\  \Gamma_{12}[\mathcal{F}] \ \ \\  
& = \ -n_c[1-e^{-\beta(\mu-\mu_c -(u-v_c)^2/2)}]\sum_{k_2,k_3\in\mathbb{N}\cup\{0\},k_1\in\mathbb{N}}\gamma^3 e^{-\frac{|v_c-u|^2(k_1+k_2+k_3)}{2\tau}}\times \\
& \times e^{\frac{(-2\mathcal{E}_c+2U+v_c^2)k_1}{2\tau}}\int_{x\cdot y = \frac{v_c^2}{2}+U-\mathcal{E}_c}e^{-(k_1+k_2)[|x|^2+x\cdot(v_c-u)]-(k_1+k_3)[|y|^2+y\cdot(v_c-u)]/(2\tau)}dxdy,
\end{aligned}
\end{equation}
with the notice that  from \eqref{gamma} and \eqref{tau}, $\gamma$ and $\tau$ are functions of ${n_n}$ and ${\mathbb{E}_n}$.

\section{The two-fluid Euler quantum hydrodynamic limit}\label{Sec:Euler}
In this section, we will derive a two-fluid Euler quantum hydrodynamic limit from \eqref{BECSuperFluid} - \eqref{RescaledSystem2a}. In this case, $\epsilon$ is a constant, so we will set it to be $\epsilon=1$. Choose $\tilde\epsilon$ to be any small parameter. 
In order to obtain the Euler hydrodynamics limit, let us start with the following Hilbert expansion using $\tilde\epsilon$ as the small parameter (cf. \cite{Caflisch:TFD:1980}):
\begin{equation}\label{Hilbert}
f \ = \ \sum_{i=0}^n \tilde\epsilon^i f^{(i)} \ + \ \tilde\epsilon^l \varsigma,
\end{equation}
in which $n$ and $l$ are positive integers. As a consequence, we can replace $f$ by its Hilbert expansion into
$$\mathcal{D}f \ = \ C_{22}[f],$$
to get a linear system of equations and a weakly nonlinear equation for the remainder $\varsigma$,  which reads as:
\begin{eqnarray}\label{HilbertSystem1}
B_1(f^{(0)},f^{(0)}) \ + \ B_2(f^{(0)},f^{(0)},f^{(0)})  & = &  0,\\\label{HilbertSystem2}
2B_1(f^{(0)},f^{(1)}) \ + \ 3B_2(f^{(0)},f^{(0)},f^{(1)})  & = & \mathcal{D}f^{(0)},\\\nonumber
2B_1(f^{(0)},f^{(i)}) \ + \ 3B_2(f^{(0)},f^{(0)},f^{(i)})  & = & \mathcal{D}f^{(i-1)}\ - \ \sum_{j=1}^{i-1}B_1(f^{(i)},f^{(i-j)})\\\label{HilbertSystem3}
& &  - \ \sum_{j,k=1, 0<j+k<i}^{i-1}B_2(f^{(i)},f^{(k)},f^{(i-j-k)}),
\end{eqnarray}
for $i=2,3,\cdots,n$. 

The equation for the remainder $r$ is as follows:
\begin{equation}\label{RemainderEquation}
\begin{aligned}
\mathcal{D}\varsigma \ =& \ \frac{1}{\tilde\epsilon}\mathcal{L}\varsigma \ + \ 2\sum_{i=1}^n\tilde\epsilon^{i-1}B_1(f^{(i)},\varsigma)\ + \ \tilde\epsilon^{l-1}B_1(\varsigma,\varsigma)\ + \ 3\sum_{i=1}^nB_2(\mathcal{F},f^{(i)},\varsigma)\\
& \ +3\sum_{i,j=1}^{n}\tilde\epsilon^{i+j-1}B_2(f^{(i)},f^{(j)},\varsigma) \ + \ 3\tilde\epsilon^{(l-1)}B_2(\mathcal{F},\varsigma,\varsigma) \ + \ 3\tilde\epsilon^{l-1}\sum_{i=1}^n \tilde\epsilon^{i}B_2(f^{(i)},\varsigma,\varsigma)\\
& \ + \tilde\epsilon^{2l-1}B_2(\varsigma,\varsigma,\varsigma) \ + \ \tilde\epsilon^{n-1}\mathfrak{Q},
\end{aligned}
\end{equation}
where $\mathfrak{Q}$ is an operator of $\mathcal{F}, f^{(1)},\cdots, f^{(n)}$.

Let us now consider each equation in the above system. From the first equation \eqref{HilbertSystem1}, we deduce that $f^{(0)} $ has to be a Bose-Einstein distribution:
\begin{equation}\label{Eulerf0}
f^{(0)} \ = \ {\mathcal{F}}.
\end{equation}
The equations \eqref{HilbertSystem2} and \eqref{HilbertSystem3} lead to linear integral equations for $f^{(1)},\cdots, f^{(i)}$. Thanks to Fredhom's theory, these linear integral equations are solvable if the right hand sides are orthogonal to $\mathcal{N}$ in $L^2(\mathbb{R}^3)$. As a consequence, $f^{(1)}$ can be solved from \eqref{HilbertSystem2}, if the following condition is satisfied
\begin{equation}\label{EulerSolvable}
\mathbb{P}\mathcal{D}{\mathcal{F}} \ = \ 0.
\end{equation}
We recall that  $\mathbb{P}$ and $\mathbb{P}^{\bot}=1-\mathbb{P}$ are the orthogonal projection operators onto $\mathcal{N}$ and $\mathcal{R}$ in $L^2(\mathbb{R}^3)$.

\subsection{The Euler quantum hydrodynamic limit of the thermal cloud kinetic equation}\label{Sec:EulerLimit}
Integrating Equation \eqref{QBFull} in $p$, we obtain
\begin{equation}\label{Sec:EulerLimit:E1}
\begin{aligned}
&\ \partial_t \int_{\mathbb{R}^3}f(t,r,p)dp \ + \ \nabla_r\cdot\int_{\mathbb{R}^3} {p}f(t,r,p)dp - \int_{\mathbb{R}^3}\nabla_r U (t,r,p)\cdot \nabla_p f(t,r,p)dp\\ 
=&\ \int_{\mathbb{R}^3}C_{12}[f](t,r,p)dp \ + \ \int_{\mathbb{R}^3}C_{22}[f](t,r,p)dp.
\end{aligned}
\end{equation}
Using the fact that 
$$\int_{\mathbb{R}^3}\nabla_r U(t,r) \cdot\nabla_p f(t,r,p)dp \ = \ \int_{\mathbb{R}^3}C_{22}[f](t,r,p)dp \ = \ 0,$$
we get
\begin{equation}\label{Sec:EulerLimit:E2}
\begin{aligned}
&\ \partial_t \int_{\mathbb{R}^3}fdp \ + \ \nabla_r\cdot\int_{\mathbb{R}^3}{p} fdp \ 
=&\ \Gamma_{12}[f].
\end{aligned}
\end{equation}
Equation \eqref{Sec:EulerLimit:E2} can be rewritten as
\begin{equation}\label{Sec:EulerLimit:MomentEq1}
\partial_t n_n \ + \ \nabla_r\cdot(n_n v_n)\
=\ \Gamma_{12}[f].
\end{equation}
For an arbitrary momentum vector $p=(p_1,p_2,p_3)$, we choose $p_j$, $j\in\{1,2,3\}$ as a test function for \eqref{QBFull} and obtain
\begin{equation}\label{Sec:EulerLimit:E4}
\begin{aligned}
&\ \partial_t \int_{\mathbb{R}^3}f(t,r,p)p_j dp \ + \ \nabla_r\cdot\int_{\mathbb{R}^3} {p}f(t,r,p)p_j dp - \int_{\mathbb{R}^3}\nabla_r U(t,r) \cdot \nabla_p f(t,r,p) p_jdp\\ 
=&\ \int_{\mathbb{R}^3}p_j C_{12}[f](t,r,p)dp \ + \ \int_{\mathbb{R}^3}p_j C_{22}[f](t,r,p)dp.
\end{aligned}
\end{equation}
Due to the conservation of momentum for $C_{12}$ and $C_{22}$, 
$$\int_{\mathbb{R}^3}(p_j - v_{cj}) C_{12}[f]dp  \ = \ \int_{\mathbb{R}^3}p_jC_{22}[f]dp \ = \ 0,$$
we get
\begin{equation}\label{Sec:EulerLimit:E5}
\begin{aligned}
&\ \partial_t \int_{\mathbb{R}^3}f(t,r,p)p_j dp \ + \ \nabla_r\cdot\int_{\mathbb{R}^3} {p}f(t,r,p)p_j dp - \int_{\mathbb{R}^3}\nabla_r U(t,r)\cdot \nabla_p f(t,r,p) p_jdp\\ 
=&\ \int_{\mathbb{R}^3}v_{cj}(t,r) C_{12}[f](t,r,p)dp\\ 
=&\ v_{cj}(t,r) \Gamma_{12}[f](t,r).
\end{aligned}
\end{equation}
Let us look at the first term on the left hand side of \eqref{Sec:EulerLimit:E5}
\begin{equation}\label{Sec:EulerLimit:E6}
\begin{aligned}
 \partial_t \int_{\mathbb{R}^3}fp_j dp \ = &\  \partial_t(n_nv_{nj})\ = \ \partial_t n_n v_{nj}\ + \ n_n\partial_t v_{nj},
\end{aligned}
\end{equation}
in which $v_{nj}$ is the component of $v_n = (v_{n1}, v_{n2}, v_{n3})$.
\\ By using \eqref{Sec:EulerLimit:MomentEq1}, we can deduce from \eqref{Sec:EulerLimit:E6} that
\begin{equation}\label{Sec:EulerLimit:E6a}
\begin{aligned}
 \partial_t \int_{\mathbb{R}^3}fp_j dp \ = &\  \Gamma_{12}[f]v_{nj}\ - \ v_{nj}\nabla_r \cdot (n_n v_n) \ + \ n_n\partial_t v_{nj},
\end{aligned}
\end{equation}
\\ Now, let us look at the second term on the left hand side of  \eqref{Sec:EulerLimit:E5}, 
\begin{equation}\label{Sec:EulerLimit:E7}
\begin{aligned}
 \nabla_r\cdot\int_{\mathbb{R}^3}{p} fp_j dp\
  = & \ \sum_{i=1}^3  \partial_{r_i}\int_{\mathbb{R}^3}{p_ip_j}fdp\\
 = & \ \sum_{i=1}^3  \partial_{r_i}\int_{\mathbb{R}^3}[{(p_i-v_{ni})(p_j-v_{nj}) + p_iv_{nj} + p_jv_{ni} -v_{ni}v_{nj}}]fdp\\
  = & \ \sum_{i=1}^3  \partial_{r_i}\int_{\mathbb{R}^3}{(p_i-v_{ni})(p_j-v_{nj})}fdp \\
  &\ + \ \sum_{i=1}^3  \partial_{r_i}\int_{\mathbb{R}^3}({p_iv_{nj} + p_jv_{ni}})fdp \ - \ \sum_{i=1}^3  \partial_{r_i}\int_{\mathbb{R}^3} v_{ni}v_{nj}fdp.
\end{aligned}
\end{equation}
By observing that 
\begin{equation}\label{Sec:EulerLimit:E8}
\begin{aligned}
 \int_{\mathbb{R}^3}({p_iv_{nj} + p_jv_{ni}})f(t,r,p)dp \ = & \ 2v_{nj}(t,r)v_{ni}(t,r)n_n(t,r)  \\
\int_{\mathbb{R}^3} v_{ni}v_{nj}f(t,r,p)dp \ = & \ v_{nj}(t,r)v_{ni}(t,r)n_n(t,r),
\end{aligned}
\end{equation}
we infer from Identity \eqref{Sec:EulerLimit:E7} 
\begin{equation}\label{Sec:EulerLimit:E9}
\begin{aligned}
 \nabla_r\cdot\int_{\mathbb{R}^3}{p} fp_j dp
  = & \ \sum_{i=1}^3  \partial_{r_i}\int_{\mathbb{R}^3}{(p_i-v_{ni})(p_j-v_{nj})}fdp \ + \ \sum_{i=1}^3 \partial_{r_i}[v_{nj}v_{ni}n_n].
\end{aligned}
\end{equation}
The last term on the left hand side of  \eqref{Sec:EulerLimit:E5} can be rewritten in the following form, by integration by parts and the definition of $n_n$
\begin{equation}\label{Sec:EulerLimit:E10}
\begin{aligned}
-\int_{\mathbb{R}^3}\nabla_r U\cdot \nabla_p fp_j dp \ = & \ \int_{\mathbb{R}^3}\partial_{r_j} U fdp
\ =  \ n_n\partial_{r_j} U.
\end{aligned}
\end{equation}
Putting the three terms \eqref{Sec:EulerLimit:E6a}, \eqref{Sec:EulerLimit:E9} and \eqref{Sec:EulerLimit:E10} together, we find
\begin{equation}\label{Sec:EulerLimit:MomentEq2}
\begin{aligned}
n_n\left(\partial_t + v_n\cdot \nabla \right) v_{nj}\ = & \ -\sum_{i=1}^3 \partial_{r_j} \mathcal{P}[f]_{ij} -n_n\partial_{r_j}U \ -(v_{nj}-v_{cj})\Gamma_{12}[f],
\end{aligned}
\end{equation}
where
\begin{equation}\label{Sec:EulerLimit:E11}
\mathcal{P}[f]_{ij} \ = \ \int_{\mathbb{R}^3} \left({p_i} -v_{ni}(t,r)\right)\left({p_j} -v_{nj}(t,r)\right)f(t,r,p)dp.
\end{equation}
 Choosing  $|p|^2$,  as a test function for \eqref{QBFull} yields
 \begin{equation}\label{Sec:EulerLimit:E12a}
\begin{aligned}
&\ \partial_t \int_{\mathbb{R}^3}f(t,r,p)|p|^2 dp \ + \ \nabla_r\cdot \int_{\mathbb{R}^3} {|p|^2p}f(t,r,p) dp - \int_{\mathbb{R}^3}\nabla_r U(t,r) \cdot\nabla_p f(t,r,p) |p|^2dp\\ 
=&\ \int_{\mathbb{R}^3}|p|^2 C_{12}[f](t,r,p)dp \ + \ \int_{\mathbb{R}^3}|p|^2 C_{22}[f](t,r,p)dp.
\end{aligned}
\end{equation}
Let us recall the conservation of energy for $C_{12}$ and $C_{22}$
$$\int_{\mathbb{R}^3}|p|^2 C_{22}[f]dp=0,$$
and
 \begin{equation*}\label{Sec:EulerLimit:E12a}
\begin{aligned}
0\ =&\ 2\int_{\mathbb{R}^3}\left(\mathcal{E}_p-\mathcal{E}_c\right)C_{12}[f]dp\ = \int_{\mathbb{R}^3}\left({|p|^2}+ {2}U- {2}\mu_c-{v_c^2}\right)C_{12}[f]dp,
\end{aligned}
\end{equation*}
which leads to 
 \begin{equation}\label{Sec:EulerLimit:E12}
\begin{aligned}
&\ \partial_t \int_{\mathbb{R}^3}f(t,r,p)|p|^2 dp \ + \ \nabla_r\cdot\int_{\mathbb{R}^3} {|p|^2p}f(t,r,p) dp - \int_{\mathbb{R}^3}\nabla_r U(t,r)\cdot \nabla_p f(t,r,p) |p|^2dp\\ 
=&\ \left(- {2}U + {2}\mu_c + {v_c^2}\right) \Gamma_{12}[f](t,r).
\end{aligned}
\end{equation}
Similar as above, we consider each term on the right and left hand sides of \eqref{Sec:EulerLimit:E12}. Let us start with the first term on the left hand side
 \begin{equation}\label{Sec:EulerLimit:E13}
\begin{aligned}
 \partial_t \int_{\mathbb{R}^3}f|p|^2 dp \
=&\  \partial_t \left(\int_{\mathbb{R}^3}f|p-v_n|^2 dp \right) \ + \ \partial_t \left(\int_{\mathbb{R}^3}f2p\cdot v_n dp \right)\\
&\ - \partial_t \left(\int_{\mathbb{R}^3}f|v_n|^2 dp \right), 
\end{aligned}
\end{equation}
where we have used the identity
\begin{equation}\label{Sec:EulerLimit:E13a}
|p-v_n|^2 \ + \ 2p\cdot v_n \ - \ |v_n|^2 \ = \ |p|^2.
\end{equation}
Since
$$\left(\int_{\mathbb{R}^3}fp\cdot v_n dp \right)\ = \ |v_n|^2n_n \ = \ \left(\int_{\mathbb{R}^3}f|v_n|^2 dp \right),$$
we obtain from \eqref{Sec:EulerLimit:E13} that
 \begin{equation}\label{Sec:EulerLimit:E14}
\begin{aligned}
\partial_t \int_{\mathbb{R}^3}f|p|^2 dp \
=&\  \partial_t \left(\int_{\mathbb{R}^3}f|p-v_n|^2 dp \right) \ + \ \partial_t \left(|v_n|^2n_n\right)\\
=&\  2\partial_t \mathbb{E} \ + \ \partial_t \left(|v_n|^2n_n\right).
\end{aligned}
\end{equation}
Expanding the second term on the right hand side of \eqref{Sec:EulerLimit:E14} gives us
 \begin{equation}\label{Sec:EulerLimit:E16}
\begin{aligned}
\partial_t \int_{\mathbb{R}^3}f|p|^2 dp \
=&\  2\partial_t \mathbb{E} \ + \ 2 n_n v_n\cdot \partial_t v_n \ + \  |v_n|^2 \partial_t n_n, 
\end{aligned}
\end{equation}
which, by \eqref{Sec:EulerLimit:MomentEq1} and \eqref{Sec:EulerLimit:MomentEq2}, can be rewritten as
 \begin{equation}\label{Sec:EulerLimit:E17}
\begin{aligned}
& \partial_t \int_{\mathbb{R}^3}f|p|^2 dp \\
=&\  2\partial_t \mathbb{E} \ + \ \sum_{j=1}^3 2 v_{nj}\left[ -\sum_{i=1}^3 \partial_{r_j} \mathcal{P}[f]_{ij} -n_n\partial_{r_j}U  \ -(v_{nj}-v_{cj})\Gamma_{12}[f]- n_nv_n\cdot \nabla_r v_{nj}\right] \\
&\  + \  |v_n|^2  [\Gamma_{12}[f] \ - \ \nabla_r\cdot(n_nv_n)]. 
\end{aligned}
\end{equation}
Now, for the second term on the left hand side of \eqref{Sec:EulerLimit:E12}, it is straight forward that
 \begin{equation}\label{Sec:EulerLimit:E18}
\begin{aligned}
 \nabla_r\cdot\int_{\mathbb{R}^3} {|p|^2p}f dp \
=&\ \nabla_r\cdot\left(\int_{\mathbb{R}^3} {|p-v_n|^2(p-v_n)}f dp\right)\ + \ \nabla_r \cdot\left(\int_{\mathbb{R}^3}|v_n|^2v_nfdp\right)\\ 
&\ -3\nabla_r\cdot \left(\int_{\mathbb{R}^3} |v_n|^2p fdp\right)\ + \ 3\nabla_r\cdot \left(\int_{\mathbb{R}^3}|p|^2v_nf dp\right),
\end{aligned}
\end{equation}
which, as a view of the identity 
$$\int_{\mathbb{R}^3}|v_n|^2v_nfdp \ = \ \int_{\mathbb{R}^3} |v_n|^2p fdp \ = \ |v_n|^2v_nn_n,$$
can  be expressed as
 \begin{equation}\label{Sec:EulerLimit:E19}
\begin{aligned}
&\ \nabla_r\cdot\int_{\mathbb{R}^3} {|p|^2p}fdp \\ 
=&\ \nabla_r\cdot\left(\int_{\mathbb{R}^3} {|p-v_n|^2(p-v_n)}fdp\right)\ - \ 2\nabla_r\cdot \left(|v_n|^2v_nn_n\right)\\ 
&\ + \ 3\nabla_r\cdot \left(\int_{\mathbb{R}^3}|p|^2v_nf dp\right).
\end{aligned}
\end{equation}
Using \eqref{Sec:EulerLimit:E13a}, we can rewrite \eqref{Sec:EulerLimit:E19} as
 \begin{equation}\label{Sec:EulerLimit:E20}
\begin{aligned}
\nabla_r\cdot\int_{\mathbb{R}^3} {|p|^2p}f dp \
=&\ \nabla_r\cdot\left(\int_{\mathbb{R}^3} {|p-v_n|^2(p-v_n)}f dp\right)\ - \ 2\nabla_r\cdot \left(|v_n|^2v_nn_n\right)\\ 
&+3\nabla_r \cdot\left(\int_{\mathbb{R}^3}|p-v_n|^2v_nf dp + 2|v_n|^2\int_{\mathbb{R}^3}pf dp -|v_n|^2v_n\int_{\mathbb{R}^3}f dp\right),
\end{aligned}
\end{equation}
which can be reduced to
 \begin{equation}\label{Sec:EulerLimit:E21}
\begin{aligned}
 \nabla_r\cdot\int_{\mathbb{R}^3} {|p|^2p}f dp \
=&\ \nabla_r\cdot\left(\int_{\mathbb{R}^3} {|p-v_n|^2(p-v_n)}f dp\right)\ - \ 2\nabla_r \cdot\left(|v_n|^2v_nn_n\right)\\ 
&+3\nabla_r\cdot \left(\int_{\mathbb{R}^3}|p-v_n|^2v_nf dp + |v_n|^2v_nn_n\right)\\ 
=&\ \nabla_r\cdot\left(\int_{\mathbb{R}^3} {|p-v_n|^2(p-v_n)}f dp\right)\ + \ \nabla_r\cdot \left(|v_n|^2v_nn_n\right)\\ 
&+3\nabla_r\cdot \left(v_n\int_{\mathbb{R}^3}|p-v_n|^2fdp\right),
\end{aligned}
\end{equation}
The last term on the left hand side of \eqref{Sec:EulerLimit:E12} can be rewritten in a straightforward manner as follows:
\begin{equation}\label{Sec:EulerLimit:E22}
\begin{aligned}
\int_{\mathbb{R}^3}|p|^2\nabla_r U \cdot\nabla_p  f dp\ 
=&\ - 2\nabla_r U \cdot \int_{\mathbb{R}^3} p  f dp.
\end{aligned}
\end{equation}
Notice that the right hand side of \eqref{Sec:EulerLimit:E22} can be expressed in terms of $n_n$ and $v_n$ as
\begin{equation}\label{Sec:EulerLimit:E23}
\begin{aligned}
- 2\nabla_r U \cdot \int_{\mathbb{R}^3} p  f dp\ 
=&\ - 2\nabla_r U \cdot\left(n_nv_n\right).
\end{aligned}
\end{equation}
As a consequence, we find
\begin{equation}\label{Sec:EulerLimit:E24}
\begin{aligned}
\int_{\mathbb{R}^3}|p|^2\nabla_r U \cdot \nabla_p fdp\ 
=&\ - 2\nabla_r U \cdot\left(n_nv_n\right).
\end{aligned}
\end{equation}
Combining \eqref{Sec:EulerLimit:E12}, \eqref{Sec:EulerLimit:E17}, \eqref{Sec:EulerLimit:E21} and \eqref{Sec:EulerLimit:E24}, yields
 \begin{equation}\label{Sec:EulerLimit:E25}
\begin{aligned}
&\  2\partial_t \mathbb{E} \ + \ \sum_{j=1}^3 2 v_{nj}\left[ -\sum_{i=1}^3 \partial_{r_j} \mathcal{P}[f]_{ij} -n_n\partial_{r_j}U  \ -(v_{nj}-v_{cj})\Gamma_{12}[f]- n_nv_n\cdot \nabla_r v_{nj}\right] \\
&\  + \ |v_n|^2  [\Gamma_{12}[f] \ - \ \nabla_r\cdot(n_nv_n)]\\ 
&\ + \ \nabla_r\cdot\left(\int_{\mathbb{R}^3} {|p-v_n|^2(p-v_n)}f dp\right)\ + \ \nabla_r \cdot\left(|v_n|^2v_nn_n\right)\\ 
&+3\nabla_r \cdot\left(v_n\int_{\mathbb{R}^3}|p-v_n|^2f dp\right) - 2\nabla_r U \cdot\left(n_nv_n\right)\\
=&\ \left(- {2}U + {2}\mu_c + {v_c^2}\right) \Gamma_{12}[f],
\end{aligned}
\end{equation}
which leads to 
 \begin{equation}\label{Sec:EulerLimit:MomentEq3}
\begin{aligned}
&\  \partial_t \mathbb{E} \ + \ \nabla_r\cdot (\mathbb{E}v_n)\\
=&\ -\nabla_r \cdot \mathcal{R}[f] -\sum_{i,j=1}^3\frac{1}{2}\left(v_{ni}\partial_{r_j}+v_{nj}\partial_{r_i}\right)P_{ij}+\left(\frac{(v_n-v_c)^2}{2}+\mu_c-U\right) \Gamma_{12}[f],
\end{aligned}
\end{equation}
where 
\begin{equation}\label{Sec:EulerLimit:E26}
\mathcal{R}[f](t,r) \ = \int_{\mathbb{R}^3}\frac{|p-v_n|^2(p-v_n)}{2}f(t,r,p)dp.
\end{equation}
The three equation \eqref{Sec:EulerLimit:MomentEq1}, \eqref{Sec:EulerLimit:MomentEq2} and \eqref{Sec:EulerLimit:MomentEq3} lead to the following system of moment equations for the kinetic equation of the thermal cloud:
\begin{equation}\label{Sec:EulerLimit:MomentEqFinal}
\begin{aligned}
\partial_t n_n \ + \ \nabla_r\cdot(n_nv_n)\
= &\ \Gamma_{12}[f],\\
n_n\left(\partial_t + v_n\cdot \nabla \right) v_{nj}\ = & \ -\sum_{i=1}^3 \partial_{r_j} \mathcal{P}[f]_{ij} -n_n\partial_{r_j}U  \ -(v_{nj}-v_{cj})\Gamma_{12}[f],\\
\  \partial_t \mathbb{E} \ + \ \nabla_r\cdot (\mathbb{E}v_n)
=&\ -\nabla_r \cdot \mathcal{R}[f] -\sum_{i,j=1}^3\frac{1}{2}\left(v_{ni}\partial_{r_j}+v_{nj}\partial_{r_i}\right)P_{ij}\\
& \ +\left(\frac{(v_n-v_c)^2}{2}+\mu_c-U\right) \Gamma_{12}[f].
\end{aligned}
\end{equation}
Replacing $\cal F$ into \eqref{Sec:EulerLimit:MomentEqFinal}, we obtain $R[\mathcal{F}]=0$, 
\begin{equation}\label{Sec:EulerLimit:E28}
P_{ij}(t,r) \ = \ \delta_{ij} \tilde{\mathbb{E}}(t,r)  \  \equiv \  \delta_{ij} \int_{\mathbb{R}^3}\frac{|p|^2}{3}f_\infty(t,r,p) dp.
\end{equation}
Moreover, we note that 
\begin{equation}\label{Sec:EulerLimit:E29}
\mathbb{E}(t,r) \ = \ \frac{3}{2}\tilde{\mathbb{E}}(t,r) .
\end{equation}
As a consequence, we can close the system \eqref{Sec:EulerLimit:MomentEqFinal} to obtain
\begin{equation}\label{Sec:EulerLimit:EulerEqbis}
\begin{aligned}
 {\partial_t n_c}\ +  \ \nabla_r\cdot(n_cv_c)\
=&\  - \ \Gamma_{12}[\mathcal{F}],\\
{n_n}\left(\partial_t + v_n\cdot \nabla \right) v_{n}\ = & \ - \nabla_r \tilde{\mathbb{E}}_n-{n_n}\nabla_r U  \ -(v_{n}-v_{c})\Gamma_{12}[\mathcal{F}],\\
\  \partial_t \tilde{\mathbb{E}}_n \ + \ \nabla_r\cdot (\tilde{\mathbb{E}}_nv_n)
=&\ -\frac{2}{3}\tilde{\mathbb{E}}_n\nabla_r \cdot v_n  \ +\frac{2}{3}\left(\frac{(v_n-v_c)^2}{2}+\mu_c-U\right) \Gamma_{12}[\mathcal{F}]
.
\end{aligned}
\end{equation}


\subsection{Comparison with a previous result}\label{Comparison}
Putting the two systems \eqref{BECSuperFluid} and \eqref{Sec:EulerLimit:EulerEqbis} together, one finds the following two-fluid Euler quantum hydrodynamic 

\begin{equation}\label{EulerTwoFluidG1}
\begin{aligned}
 {\partial_t n_c}\ +  \ \nabla_r\cdot(n_cv_c)\
=&\  - \ \Gamma_{12}[\mathcal{F}],
\\
\partial_t v_c +\frac{\nabla_r v_c^2}{2} = & \ -\nabla_r\mu_c,\\
\partial_t {n_n} \ + \ \nabla_r\cdot({n_n}v_n)
= &\ \Gamma_{12}[\mathcal{F}],\\
{n_n}\left(\partial_t + v_n\cdot \nabla \right) v_{nj}\ = & \ - \nabla_r \tilde{\mathbb{E}}_n-{n_n}\nabla_r U  \ - (v_{nj}-v_{cj})\Gamma_{12}[\mathcal{F}],\\
\  \partial_t \tilde{\mathbb{E}}_n \ + \ \nabla_r\cdot (\tilde{\mathbb{E}}_nv_n)
=&\ -\frac{2}{3}\tilde{\mathbb{E}}_n\nabla_r \cdot v_n +\frac{2}{3}\left(\frac{(v_n-v_c)^2}{2}+\mu_c-U\right) \Gamma_{12}[\mathcal{F}].
\end{aligned}
\end{equation}

In \eqref{EulerTwoFluidG1}, the condensate and non-condensate parts are coupled through both $\mu_c$ and $\Gamma_{12}$. Notice that $\Gamma_{12}$ is already computed in \eqref{Gamma12a} and \eqref{Gamma12b}.


In the thesis \cite{Allemand:Thesis:2009}, the author has derived the following hydrodynamic limit:

\begin{equation}\label{EulerTwoFluideG1}
\begin{aligned}
 {\partial_t n_c}\ +  \ \nabla_r\cdot(n_cv_c)& \
=\  0,\\
\partial_t {n_n} \ + \ \nabla_r\cdot({n_n}v_n)&\ 
= \ 0,\\
\partial_t(n_nv_n) \ + \ \nabla_r\cdot(n_nv_n \otimes	 v_n + \mathbb{E}_n I_3)& \ = \ - gn_n\nabla_r\cdot(2n_n+n_c),\\
\partial_t(n_cv_c) \ + \ \nabla_r\cdot(n_cv_c \otimes	 v_c)& \ = \ - \frac{g}{2}n_c\nabla_r\cdot(2n_n+n_c),\\
\partial_t(\frac12 n_n|v_n|^2+\frac12 n_c|v_c|^2+\frac32\mathbb{E}_n +\frac{g}{4}(2n_n+n_c)^2)) \ & \\
+ \nabla_r\cdot\Big(\frac12 n_n|v_n|^2v_n+\frac12 n_c|v_c|^2v_c+\frac52\mathbb{E}_n v_n  \ & \\
+ \frac{g}{2}(2n_n+n_c)(2n_nv_n+n_cv_c) \Big) & \ =   0,
\end{aligned}
\end{equation}
where $I_3$ is the identity $3\times3$ matrix. It is also mentioned \cite{Allemand:Thesis:2009} that this is a  two-phases Euler system, the second fluid (the superfluid) being pressureless and they do not exchange mass, contrary
to what occurs the Landau two-fluid theory \cite{bogoliubov1970lectures,lifshitz1987fluid}.

On the contrary, our limit \eqref{EulerTwoFluidG1} agree with the Landau two-fluid theory \cite{bogoliubov1970lectures,lifshitz1987fluid}. The main reason is that, following \cite{GriffinNikuniZaremba:2009:BCG}, in general  excited atoms in the condensate need not to be in local  equilibrium with the condensate atoms. As a consequence, $C_{12}$ and $C_{22}$, in most of the cases, do not share the same equilibrium distribution. Our equilibrium distribution $\mathcal{F}$ is the natural equilibrium used in most physical contexts \cite{GriffinNikuniZaremba:2009:BCG} and $C_{22}[\mathcal{F}]=0$ but $C_{12}[\mathcal{F}]\ne 0$. Therefore, the two fluids are coupled.

In \cite{Allemand:Thesis:2009}, the author considers a very special choice of $\mathcal{F}$ 
$$\mathcal{F}(t,r,p) \ = \ \frac{1}{e^{\beta[(p-v_n)^2/2 - |v_c-v_n|^2/2-U/2]}-1},
$$
where the temperature parameter $\beta$ is a constant, instead of being a function of $(t,r)$. Moreover, the  effect of the chemical potential $\mu(t,r)$ is also ignored. This special choice of the distribution  $\mathcal{F}$ implies $C_{22}[\mathcal{F}]=C_{12}[\mathcal{F}]=0$. The two fluids are then decoupled, that is in contradiction with the  Landau two-fluid theory \cite{bogoliubov1970lectures,lifshitz1987fluid}, as the author pointed out.

\section{The two-fluid Navier-Stokes quantum hydrodynamic approximations}\label{Sec:NavierStokes}

This section is devoted to the derivation of the Navier-Stokes approximation of the system \eqref{RescaledSystem2b}  - \eqref{RescaledSystem2a} through the Chapman-Enskog expansion, under the assumption $g=\epsilon^{\delta_0}$.  Similar as in Section \ref{Sec:EulerLimit}, we also have the   expansion:
\begin{equation}\label{Sec:NavierStokes:E3}
f \ = \ \sum_{i=0}^n \epsilon^i f^{(i)} \ + \ \epsilon^l \varsigma,
\end{equation}
in which $n$ and $l$ are positive integers.

Arguing similarly as above, we deduce that $f^{(0)} $ has to be a Bose-Einstein distribution:
\begin{equation}\label{Eulerf0q}
f^{(0)} \ = \ {\mathcal{F}}.
\end{equation}

Decompose $f^{(i)}$ into two parts
\begin{equation}\label{Sec:NavierStokes:E8}
f^{(i)} \ = \ h^{(i)} \ + \ k^{(i)},
\end{equation}
where
$$h^{(i)} \ \in \mathcal{R}, \ \ \ k^{(i)} \ \in \mathcal{N}.
$$
From \eqref{HilbertSystem2}, one has
\begin{equation}\label{Sec:NavierStokes:E9}
h^{(1)}\ = \ \mathcal{L}^{-1}\mathcal{D}\mathcal{F}.
\end{equation}

Adopting  the same techniques used in \cite{ArlottiLachowicz:EAN:1997,KawashimaMatsumuraNishida:OTF:1979}, we decompose $h^{(1)}$ into the sum of $h'$ and $h''$:
$$h^{(1)} \ = \ h' \ + \ h'',$$
where $h'$ and $h''$ satisfy the following system of equations:
\begin{eqnarray}\label{Sec:NavierStokes:E10}
\mathcal{L}h' & = & \mathbb{P}^{\bot}\mathcal{D}\mathcal{F},\\\label{Sec:NavierStokes:E11}
\mathbb{P}\mathcal{D}\mathcal{F} & = & -\epsilon \mathbb{P}\mathcal{D}h',\\\label{Sec:NavierStokes:E12}
\mathcal{L}h'' & = & \epsilon\mathbb{P}^{\bot}\mathcal{D}k^{(1)},\\\label{Sec:NavierStokes:E13}
\mathbb{P}\mathcal{D}k^{(1)} & = & -\mathbb{P}\mathcal{D}h'',
\end{eqnarray}
and
\begin{equation}\label{Sec:NavierStokes:E14}
\mathcal{L}h^{(i)} \ = \ \epsilon\mathbb{P}^{\bot}\mathcal{D}k^{(i)} \ + \ \mathbb{P}^{\bot}\mathcal{D}h^{(i-1)}\ - \ \sum_{j=1}^{i-1}{Q}_1(f^{(j)},f^{(i-j)}) \ - \ \sum_{j,k=0,0<j+k<i}^{i-1}{Q}_2(f^{(j)},f^{(k)},f^{(i-j-k)}),
\end{equation}
\begin{equation}\label{Sec:NavierStokes:E15}
\mathbb{P}\mathcal{D}k^{(i)}\ = \ - \mathbb{P}\mathcal{D}h^{(i)}.
\end{equation}

By the Fredholm theory, the system \eqref{Sec:NavierStokes:E10}-\eqref{Sec:NavierStokes:E13} can be solved in $L^2(\mathbb{R}^3)$, if 
\begin{equation}\label{Sec:NavierStokes:E17}
\begin{aligned}
h'  \ = & \ \mathcal{L}^{-1}(\mathbb{P}^\bot \mathcal{D}\mathcal{F}),\\
h''  \ = & \ \mathcal{L}^{-1}(\epsilon\mathbb{P}^\bot \mathcal{D}k^{(1)})
\end{aligned}
\end{equation}
and 
\begin{equation}\label{Sec:NavierStokes:E18}
\begin{aligned}
h^{(i)}\ = & \ \mathcal{L}^{-1}\Big(\epsilon\mathbb{P}^{\bot}\mathcal{D}k^{(i)}\ + \ \mathbb{P}^\bot \mathcal{D}h^{(i-1)}\ - \ \sum_{j=1}^{i-1}{B}_1(f^{(j)},f^{(i-j)})\\
& \  - \ \sum_{j,k=0;0<j+k<i}^{i-1}{B}_2(f^{(j)},f^{(k)},f^{(i-j-k)})\Big),
\end{aligned}
\end{equation}
for $i=2,3,\cdots$

The  equation \eqref{Sec:NavierStokes:E17} yields
\begin{equation}\label{Sec:NavierStokes:E19}
\mathbb{P}\mathcal{D}\mathcal{F} \ = \ -\epsilon \mathbb{P}\mathcal{D}h'\ = \ -\epsilon \mathbb{P}\mathcal{D}\mathcal{L}^{-1}\mathbb{P}^\bot\mathcal{D}\mathcal{F}.
\end{equation}

Equation \eqref{Sec:NavierStokes:E19} leads to the Navier-Stokes approximation, which will be computed in Section \ref{Sec:NavierStokesLimit}.

\subsection{Inversion of the linearized operator of $C_{22}$}
 
Define
\begin{equation}\label{Symmetries1}
\begin{aligned}
\mathcal{A}(p)=p \otimes p -\frac{1}{3}|p|^2Id, \ \ \ \ \ \mathcal{B}(p)=\frac{1}{2}p(|p|^2-5),
\end{aligned}
\end{equation}
clearly, 
\begin{equation}\label{Orthogonal}
\mathcal{A}_{jk}\perp \mathrm{ker}\mathcal{L}, \ \  \ \ \mathcal{B}_{l}\perp \mathrm{ker}\mathcal{L}, \ \ \ \  \mathcal{B}_{l}\perp \mathcal{A}_{jk}, \ \ \ \  j,k,l =1,2,3.
\end{equation}

By the same algebraic argument as the one used for the classical Boltzmann collision operator (cf. page 64-65 \cite{BouchutGolse:2000:KEA}), one can deduce that there exists scalar-valued functions $\alpha_0(|p|)$, $\alpha_1(|p|)$ such that
\begin{equation}\label{Symmetries2}
\begin{aligned}
\mathcal{L}^{-1}\left(\frac{\mathcal{F}^2(p)}{\mathcal{M}(p)}\mathcal{A}(p)\right)= \alpha_0(|p|)\frac{\mathcal{F}^2(p)}{\mathcal{M}(p)}\mathcal{A}(p), \ \ \ \ \ \mathcal{L}^{-1}\left(\frac{\mathcal{F}^2(p)}{\mathcal{M}(p)}\mathcal{B}(p)\right)= \alpha_1(|p|)\frac{\mathcal{F}^2(p)}{\mathcal{M}(p)}\mathcal{B}(p).
\end{aligned}
\end{equation}

A direct consequence of \eqref{Symmetries2} is the existence of scalar-valued functions $\beta_0(|p|)$ and $\beta_1(|p|)$ such that

\begin{equation}\label{Symmetries3}
\begin{aligned}
\mathcal{L}^{-1}\left(\frac{\mathcal{F}^2(p)}{\mathcal{M}(p)}p^ip^j\right)\ = &\ \beta_0(|p|)\frac{\mathcal{F}^2(p)}{\mathcal{M}(p)}\mathcal{A}_{ij}(p), \\  \ \mathcal{L}^{-1}\left(\frac{\mathcal{F}^2(p)}{\mathcal{M}(p)}\left(|p|^2-\frac{10\tau\Omega_2(\gamma)}{\Omega_1(\gamma)}\right)p^i\right)\ = &\ \beta_1(|p|)\frac{\mathcal{F}^2(p)}{\mathcal{M}(p)}\mathcal{B}_i(p),
\end{aligned}
\end{equation}
where $p^i$, $\mathcal{B}_i(p)$ are the $i$-th component of the vectors $p$ and $\mathcal{B}(p)$ respectively. In addition, $\mathcal{A}_{ij}(p)$ is the $(i,j)$-th element of the matrix $\mathcal{A}(p)$. Note that these symmetry invariances are very similar to the ones obtained in the context of the classical Boltzmann collision operator (cf. Equation $(2.100)$, page 64-65 \cite{BouchutGolse:2000:KEA}); we then denote
\begin{equation}\label{LinearizedInverse}
\begin{aligned}
\mathfrak{C}_{ij}(p):= \beta_0(|p|)\frac{\mathcal{F}^2(p)}{\mathcal{M}(p)}\mathcal{A}_{ij}(p), \ \ \ \ \ \mathfrak{C}_{i}(p):= \beta_1(|p|)\frac{\mathcal{F}^2(p)}{\mathcal{M}(p)}\mathcal{B}_i(p).
\end{aligned}
\end{equation}

\subsection{Navier-Stokes quantum hydrodynamic approximation of the thermal cloud}\label{Sec:NavierStokesLimit}
In this subsection, we will derive the Navier-Stokes system resulting from \eqref{Sec:NavierStokes:E19}. First, observe that
\begin{equation}\label{Sec:NavierStokes:E20}
\begin{aligned}
\mathbb{P}^\bot\Pi\mathcal{F} \ = & \ \frac{\mathcal{F}^2}{\mathcal{M}}\sum_{i,j=1}^3 \left\{(p^i-{v_n}_i)(p^j-{v_n}_j) \ - \ \frac{1}{3}|p-{v_n}|^2\delta_{i,j}\right\}\frac{1}{\tau}\frac{\partial {v_n}_j}{\partial x_i}\\
 &\ + \ \frac{\mathcal{F}^2}{\mathcal{M}}\left\{|p-{v_n}|^2 \ - \ \frac{10\tau \Omega_2(\gamma)}{3\Omega_1(\gamma)}\right\}\sum_{i=1}^3(p^i-{v_n}_i)\frac{1}{2\tau^2}\frac{\partial \tau}{\partial r_i}.
\end{aligned}
\end{equation}

Classical techniques for the classical Boltzmann collision operator can be applied (cf. \cite{TruesdellMuncaster:FOM:1980} - pp. 456-457 and \cite{KawashimaMatsumuraNishida:OTF:1979,ArlottiLachowicz:EAN:1997}), to get
\begin{equation}\label{Sec:NavierStokes:E21}
\begin{aligned}
-\mathbb{P}\Pi\mathcal{L}^{-1}\mathbb{P}^\bot\Pi\mathcal{F} \ = & \ \frac{\mathcal{F}^2}{\mathcal{M}}\left(\sum_{k=1}^3\frac{\psi_k}{\omega_k}\left(\sum_{i=1}^3\frac{\partial}{\partial r_i}\left(\varpi(\gamma,\tau)\left(\frac{\partial {v_n}_k}{\partial r_i}+\frac{\partial {v_n}_i}{\partial r_k}\right)\right)\right)\right.\\
 &\ -\frac{2}{3}\frac{\partial}{\partial r_k}\left.\left(\varpi(\gamma,\tau)\sum_{i=1}^3\frac{\partial {v_n}_i}{\partial r_i}\right)\right) \ + \ 2\frac{\psi_4}{\omega_4}\left(\sum_{i=1}^3\frac{\partial}{\partial r_i}\left(\varrho(\gamma,\tau)\frac{\partial \tau}{\partial r_i}\right)\right.\\
 &\ -\frac{2}{3}\varrho(\gamma,\tau)\left(\sum_{i=1}^3\frac{\partial {v_n}_i}{\partial r_i}\right)^2 \ + \ \varpi(\gamma,\tau)\sum_{i,k=1}^3\frac{\partial {v_n}_k}{\partial r_i}\left.\left.\left(\frac{\partial {v_n}_k}{\partial r_i}+\frac{\partial {v_n}_i}{\partial r_k}\right)\right)\right),
\end{aligned}
\end{equation}
where 
\begin{equation}\label{Sec:NavierStokes:E22}
\varpi(\gamma,\tau) \ = \ -\frac{1}{\tau} \int_{\mathbb{R}^3}\xi_1\xi_2\mathfrak{C}_{12}(\xi)d\xi,
\end{equation}
\begin{equation}\label{Sec:NavierStokes:E23}
\varrho(\gamma,\tau) \ = \ -\frac{1}{4\tau^2} \int_{\mathbb{R}^3}|\xi|^2\xi_1\mathfrak{C}_{1}(\xi)d\xi,
\end{equation}
with $\xi_1$, $\xi_2$ are the components of the vectors $\xi=(\xi_1,\xi_2,\xi_3)$ and $\mathfrak{C}_1$, $\mathfrak{C}_{12}$ are defined in \eqref{LinearizedInverse}.

Notice that 
$$\mathcal{D}\mathcal{F}=\Pi \mathcal{F}+O(\epsilon^{\delta_0}).$$

The first order approximation in terms of $\epsilon$ of the quantity $\epsilon \mathbb{P}\mathcal{D}\mathcal{L}^{-1}\mathbb{P}^\bot\mathcal{D}\mathcal{F}$ is then $\epsilon \mathbb{P}\Pi\mathcal{L}^{-1}\mathbb{P}^\bot\Pi\mathcal{F}$. The Navier-Stokes system \eqref{Sec:NavierStokes:E19} becomes
\begin{equation}\label{Sec:NavierStokes:E24}
\mathbb{P}\mathcal{D}\mathcal{F} \ = \ -\epsilon \mathbb{P}\Pi\mathcal{L}^{-1}\mathbb{P}^\bot\Pi\mathcal{F},
\end{equation}
which, thanks to the identity \eqref{Sec:NavierStokes:E21}, leads to
\begin{equation}\label{Sec:EulerLimit:NavierStokesEq}
\begin{aligned}
\partial_t {n_n} \ + \ \nabla_r\cdot({n_n}v_n)\
= &\ \epsilon{\Gamma}_{12}[\mathcal{F} ],\\
{n_n}\left(\partial_t + v_n\cdot \nabla \right) v_{nj} \ +  \partial_{r_j} ({n_n}e_n)\ = & \ -{n_n}\nabla_r \epsilon^{\delta_0}{U}  \ -(v_{nj}-v_{cj})\epsilon{\Gamma}_{12}[\mathcal{F} ]\\
& + \ \epsilon\left[\sum_{i=1}^3\frac{\partial}{\partial r_i}\left(\bar{\varpi}({n_n},e_n)\left(\frac{\partial {v_n}_j}{r_i} + \frac{\partial {v_n}_i}{r_j}\right)\right)\right.\\
& - \ \left.\frac{2}{3}\frac{\partial}{\partial r_j}\left(\bar{\varpi}({n_n},e_n)\sum_{i=1}^3\frac{\partial {v_n}_i}{r_i}\right)\right],\\
\  \partial_t e_n \ + \ \nabla_r\cdot (e_nv_n) \ + \ \frac{2}{3}e_n\nabla_r \cdot v_n
=&\ \frac{1}{{n_n}}\Big[ \frac{2}{3}\left(\frac{(v_n-v_c)^2}{2}+\mu_c-\epsilon^{\delta_0}{U}+e_n\right) \epsilon{\Gamma}_{12}[\mathcal{F} ]\Big]\\
& \ + \frac{\epsilon}{\mathcal{G}({n_n},e_n)}\left[\sum_{i=1}^3\frac{\partial}{\partial r_i}\left(\varrho_1({n_n},e_n)\frac{\partial e_n}{\partial r_i}+\varrho_2({n_n},e_n)\frac{\partial {n_n}}{\partial r_i}\right)\right.\\
&\ +\bar{\varpi}({n_n},e_n)\sum_{i,k=1}^3\frac{\partial {v_n}_k}{\partial x_i}\left(\frac{\partial {v_n}_k}{\partial x_i}\right.+ \left.\frac{\partial {v_n}_i}{\partial x_k}\right)\\
&\  \left. -\frac{2}{3}\varpi({n_n},e_n)\left(\sum_{i=1}^3\frac{\partial {v_n}_i}{\partial x_i}\right)^2\right],
\end{aligned}
\end{equation}
where
\begin{equation}\label{Sec:EulerLimit:NavierStokesEqConstants}
\begin{aligned}
\bar{\varpi}({n_n},e_n) \  =  & \ \varpi(\gamma,\tau),\\
\mathcal{G}({n_n},e_n) \ = & \ 2^{5/2}\pi \tau^{3/2}\gamma \Omega_1(\gamma),\\  
\varrho_1({n_n},e_n) \ = & \ \varrho(\gamma,\tau)\frac{\partial \tau}{\partial e_n},\\
\varrho_2({n_n},e_n) \ = & \ \varrho(\gamma,\tau)\frac{\partial \tau}{\partial {n_n}}.
\end{aligned}
\end{equation}

Combining \eqref{BECSuperFluid}  and \eqref{Sec:EulerLimit:NavierStokesEq}, we get the ``closed system''.

Moreover, 
The Navier-Stokes system of the excitations is very different from the  Navier-Stokes system obtained from the classical Boltzmann equation (cf. \cite{TruesdellMuncaster:FOM:1980}) in several points:
\begin{itemize}
\item First, in the classical Navier-Stokes system, the viscosity coefficient $\bar{\varpi}$ and the heat conduction coefficient $\varrho_1$ depend only on $e_n$. In the above quantum Boltzmann system, they depend on both $e_n$ and $n_n$. 
\item Second, different from the classical Navier-Stokes system, the second derivatives of $n_n$ also appear in the system. 
\item Third, the Navier-Stokes system of the excitations is coupled with the system of the BEC super fluid via the quantity $\epsilon{\Gamma}_{12}[\mathcal{F} ]$, computed in \eqref{Gamma12b}. 

\end{itemize}

The Navier-Stokes system for the excitations therefore has a completely different nature in comparison with the classical Navier-Stokes equation. And, hence, one could expect more complicated behaviors, that would be a subject of our future studies. 

~~ \\{\bf Acknowledgements.} This research  was supported by NSF grants DMS-1522184 and DMS-1107291: RNMS KI-Net, by NSFC grant No. 91330203, and by the Office
of the Vice Chancellor for Research and Graduate Education at the University of Wisconsin–Madison with funding from the Wisconsin Alumni Research
Foundation.
M.-B. Tran was supported partially by ERC Advanced Grant DYCON and NSF grant DMS-1814149. Tran would like to thank Prof. Linda Reichl and Prof. Jay Robert Dorfman for the discussions on the topic. The authors would also to express gratitude to the referee for very fruitful comments and suggestions that help to improve the quality of the paper.
\bibliographystyle{plain}

 \bibliography{QuantumBoltzmann}

\end{document}